\documentclass[aps,prl,superscriptaddress,twocolumn,longbibliography]{revtex4-1}
\usepackage{bbm}
\usepackage{graphicx}
\usepackage{dcolumn}
\usepackage{bm}
\usepackage{subfigure}
\usepackage{amsmath}
\usepackage{feynmf}
\usepackage{hyperref}

\usepackage{attachfile}

\newcommand{\bk}{\boldsymbol k}

\usepackage{times}

\begin{document}

\title{Boundary Topological Superconductors}

\author{Bo-Xuan Li}
\affiliation{School of Physics, Sun Yat-Sen University, Guangzhou 510275, China}
\author{Zhongbo Yan}
\email{yanzhb5@mail.sysu.edu.cn}
\affiliation{School of Physics, Sun Yat-Sen University, Guangzhou 510275, China}

\date{\today}

\begin{abstract}
For strongly anisotropic time-reversal invariant (TRI) insulators in two and three dimensions, the band inversion can
occur respectively at all TRI momenta of a high symmetry axis and
plane. Although these classes of materials are topologically trivial as the strong and
weak $Z_{2}$ indices are all trivial, they can host an even
number of unprotected helical gapless edge states or surface Dirac cones on
some boundaries. We show in this work
that when the gapless boundary states are gapped by $s_{\pm}$-wave superconductivity,
a boundary time-reversal invariant topological superconductor (BTRITSC) characterized by a $Z_{2}$ invariant
can be realized on the corresponding boundary. Since the dimension of the BTRITSC is lower than the bulk by one,
the whole system is a second-order TRI topological superconductor.  When the boundary of the BTRITSC
is further cut open, Majorana Kramers pairs and helical gapless Majorana modes will respectively appear at the corners and
hinges of the considered sample in two and three dimensions.  Furthermore, a magnetic field can gap the
helical Majorana hinge modes of the three-dimensional second-order TRI topological superconductor 
and lead to the realization of a third-order topological superconductor with Majorana corner modes.
Our proposal can potentially be realized in insulator-superconductor
heterostructures and iron-based superconductors whose normal states take the desired inverted band structures.
\end{abstract}

\maketitle

Topological insulators (TIs) and topological superconductors (TSCs) are two classes of materials which have
a nontrivial gapped band structure in the bulk and novel gapless excitations on the boundary\cite{hasan2010,qi2011}.
The band topology of TIs with time-reversal symmetry is known to be characterized by a $Z_{2}$ invariant $\nu$
in two dimensions (2D) and four $Z_{2}$ invariants $(\nu_{0};\nu_{1}\nu_{2}\nu_{3})$ in 3D\cite{Kane2005b,fu2007b,moore2007,roy2009,fu2007a}. If the inversion symmetry
is also preserved, these $Z_{2}$ invariants can be simply inferred from the parity eigenvalues at time-reversal invariant (TRI)
momenta\cite{fu2007a}, or equivalently, the number and distribution of TRI momenta at which the band inversion occurs.
In 2D, when the band inversion occurs at an odd number of TRI momenta, $\nu$ is necessitated to take the nontrivial value,
and a TI with an odd number of helical gapless states on each edge is realized\cite{Konig2007,Knez2011,Wu2018QSHE}. Similarly, when
the band inversion occurs at an odd number of TRI momenta in 3D, the strong $Z_{2}$ index $\nu_{0}$ is also
necessitated to take the nontrivial value, and a strong TI with an odd number of Dirac cones on each surface is realized\cite{Zhang2009bise,Chen2009bise,Xia2009bise,Hsieh2009bise}.
In contrast, when the band inversion occurs at an even number of TRI momenta, the strong $Z_{2}$ index is trivial,
but some of the three weak indices $(\nu_{1}\nu_{2}\nu_{3})$ can still be nontrivial. For instance,
when the band inversion occurs at $2$ (mod $4$) TRI momenta,  at least one of $(\nu_{1}\nu_{2}\nu_{3})$
must be nontrivial according to their definition\cite{fu2007a}, leading to the realization of a weak TI in which
the surface Dirac cones only appear selectively on certain surfaces and their number  on each surface is even rather than odd\cite{Liu2016weak,Noguchi2019}.

Because of the time-reversal symmetry, the gapless boundary states in TIs are spin-momentum-locked\cite{Hsieh2009helical}.
Remarkably, this property allows the establishment of
a direct connection between TIs and TSCs. In the pioneering work of Fu and Kane\cite{fu2008},
it was shown that
when the gapless surface Dirac cones of a strong
TI are gapped  by $s$-wave superconductivity, topological superconductivity
can be realized in the $\pi$-flux  vortices, manifested by the presence of Majorana
zero modes (MZMs). The vortex MZMs also follow a $Z_{2}$ classification, which
means that there exists one topologically protected MZM only when the number of MZMs in
a $\pi$-flux vortex is odd\cite{Hosur2011MZM,Qin2019vortex2,Yan2020vortex}. On the other hand, the number of MZMs is directly connected to
the number of surface Dirac cones, therefore,
insulators with an even number of band inversions
are disfavored in this scenario.
Apparently, this rules out a considerable amount of strongly anisotropic materials which generally favor an even number of band inversions.

In this work, we build a new scenario which favors such strongly anisotropic band-inverted insulators.
Concretely, we consider insulators with both time-reversal symmetry and inversion symmetry, whose
band inversions occur at all TRI momenta
of a high symmetry axis in 2D and a high symmetry plane in 3D.
Although such band-inverted insulators are topologically trivial as the strong and
weak $Z_{2}$ indices are all trivial, they can host an even
number of unprotected helical gapless edge states in 2D and surface Dirac cones in 3D on
some boundaries. The gapless boundary states are found to form
 floating bands which are non-degenerate due to the breaking
of inversion symmetry on the boundary. By introducing $s_{\pm}$-wave superconductivity
rather than $s$-wave superconductivity to gap
the gapless floating bands,
we find that a TRI TSC characterized by a $Z_{2}$
invariant can be realized on the corresponding boundary, even though the band topology of the bulk
is necessitated to be trivial. As the TRI TSC is realized on the boundary, we term it
boundary time-reversal invariant topological superconductor (BTRITSC). Because the dimension of the BTRITSC is lower than the bulk by one,
the whole system is a second-order TRI TSC which harbors Majorana Kramers pairs (two MZMs related by time-reversal symmetry) at the sample corners in 2D and
helical gapless Majorana modes at the sample hinges in 3D\cite{
Yan2018hosc,Wang2018hosc,Zhang2019hinge,Gray2019helical,Hsu2020hosc,vu2020timereversalinvariant,wu2020boundaryobstructeda}. Remarkably,
a magnetic field can gap the
helical Majorana hinge modes in 3D and lead to the realization of a third-order TSC with Majorana corner modes\cite{Yan2019hosca,Ahn2020hosc}. The new scenario thus unveils a new route for
the realization of higher-order TSCs\cite{Langbehn2017,Shapourian2018SOTSC,Khalaf2018,Geier2018,
Zhu2018hosc,Wangyuxuan2018hosc,Hsu2018hosc,Liu2018hosc,Wuzhigang2019hosc,Volpez2019SOTSC,
Zhang2019hoscb,Wu2019hoscb,Zeng2019mcm,Bultinck2019,Ghorashi2019,Peng2019hinge,Zhu2019mixed,
Laubscher2019hosc,Pan2019SOTSC,Yan2019hoscb,Franca2019SOTSC,Majid2020hosca,
Zhang2020SOTSC,Bitan2020hosc,Wu2020SOTSC,Majid2020hoscb,wu2020boundaryobstructedb,tiwari2020chiral,Ono2020SI,ono2020enriched}.

{\it BTRITSCs in 2D.--- } To illustrate the essential physics, we start with a simple
2D Bogoliubov-de Gennes (BdG) Hamiltonian $H=\frac{1}{2}\sum_{\bk}\psi_{\bk}^{\dag}H_{\rm BdG}(\bk)\psi_{\bk}$,
where
\begin{eqnarray}
H_{\rm BdG}(\bk)&=&\left(
                     \begin{array}{cc}
                       H_{0}(\bk)-\mu\Sigma_{00} & -i\Sigma_{20}\Delta(\bk) \\
                       i\Sigma_{20}\Delta(\bk) & \mu\Sigma_{00}-H_{0}^{*}(-\bk) \\
                     \end{array}
                   \right),\nonumber\\
H_{0}(\bk)&=&\epsilon_{\rm 2D}(\bk)\Sigma_{00}+m_{\rm 2D}(\bk)\Sigma_{03}
+\lambda_{1}\sin k_{1}\Sigma_{31}\nonumber\\
&&+\lambda_{2}\sin k_{2}\Sigma_{02},\nonumber\\
\Delta(\bk)&=&\Delta_{0}-\Delta_{1}(\cos k_{1}+\cos k_{2}),\label{two}
\end{eqnarray}
and $\psi_{\bk}=(c_{\bk,a,\uparrow}, c_{\bk,b,\uparrow}, c_{\bk,a,\downarrow}, c_{\bk,b,\downarrow},
c_{-\bk,a,\uparrow}^{\dag}, c_{-\bk,b,\uparrow}^{\dag}, c_{-\bk,a,\downarrow}^{\dag}, \\c_{-\bk,b,\downarrow}^{\dag})^{T}$.
In the BdG Hamiltonian (\ref{two}), $\Sigma_{ij}=s_{i}\otimes \sigma_{j}$, where
the Pauli matrices $s_{i}$ and $\sigma_{j}$ act respectively
on the spin $(\uparrow,\downarrow)$ and orbital $(a,b)$ degrees of freedom, and $s_{0}$ and $\sigma_{0}$ are
$2\times2$ unit matrices. $H_{0}(\bk)$ describes
the normal state, with the first term characterizing the asymmetry of the conduction and valence bands,
the second term characterizing the band inversion, and the last two terms representing
spin-orbit coupling. Here we take $\epsilon_{\rm 2D}(\bk)=\epsilon_{1}\cos k_{1}+\epsilon_{2}\cos k_{2}$,
$m_{\rm 2D}(\bk)=m_{0}-m_{1}\cos k_{1}-m_{2}\cos k_{2}$,
and $\lambda_{1,2}$ are assumed to be positive. $\Delta(\bk)$ describes
the superconducting order parameter. In this work, we consider $s_{\pm}$-wave superconductivity
which can be achieved intrinsically\cite{Wang2015iFeSC,Wu2016iron,Xu2016FeSC,zhang2018iron,wang2018evidence,
kong2019half,machida2019zero,Liu2018MZM,chen2019quantized,Zhu2019MZM} or extrinsically by superconducting proximity effect from an iron-based superconductor\cite{Mazin2008,Wang2011iron}. Moreover,
the lattice constants are set to unity throughout for notational simplicity.

\begin{figure}
\subfigure{\includegraphics[width=9cm, height=8cm]{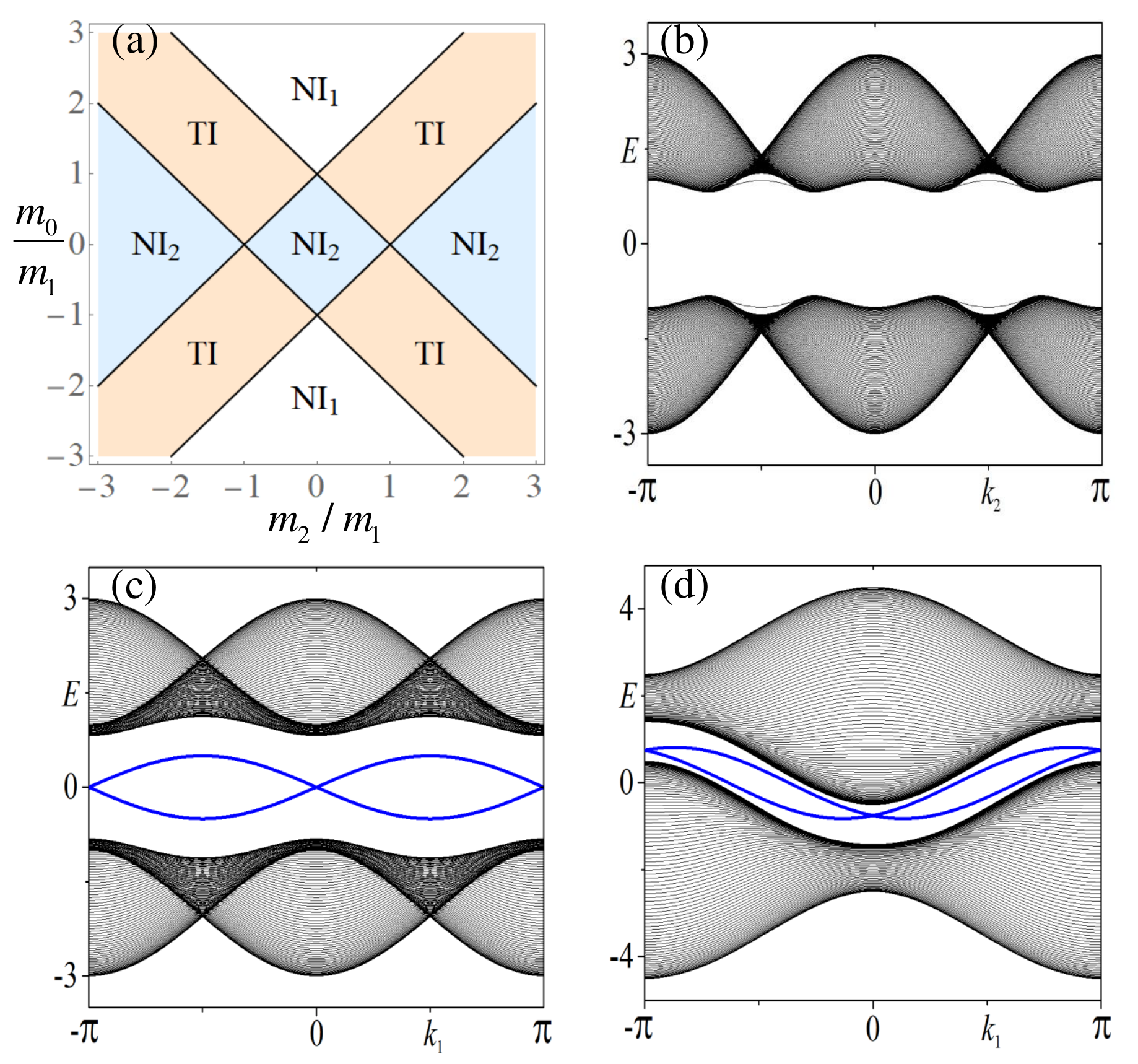}}
\caption{ (a) Phase diagram of the normal state.
 (b)(c)(d) Energy spectrum under a cylinder geometry.
 The lattice size along the direction with
open boundary condition is $L=100$. (b) No gapless state
 appears on the $x_{1}$-normal edges. (c) The in-gap dispersions (blue lines) are of double degeneracy, corresponding to
 the presence of two pairs of helical gapless states on each $x_{2}$-normal edge. The helical gapless states on
 each edge form two non-degenerate floating bands.
(d) The floating bands remain in the gap even when the conduction-valence asymmetry is strong enough
to change the insulator to a metal.
Common parameters are
$m_{1}=1$, $m_{0}=0$, $m_{2}=2$, $\lambda_{1}=0.5$, $\lambda_{2}=1$. $\epsilon_{1,2}=0$ in (b) (c),
and $\epsilon_{1}=0$, $\epsilon_{2}=1.5$ in (d).
}  \label{phase}
\end{figure}

It is readily verified that $H_{0}$ has both time-reversal symmetry and
inversion symmetry, with the time-reversal and inversion operators given by
$T=is_{2}\otimes\sigma_{0}\mathcal{K}$ and $P=s_{0}\otimes\sigma_{3}$, respectively, where
$\mathcal{K}$ denotes the complex conjugation. Accordingly, the bulk bands have Kramers  degeneracy at
every momentum,
and the $Z_{2}$ invariant characterizing the
band topology of $H_{0}(\bk)$ is simply given by\cite{fu2007a}
\begin{eqnarray}
(-1)^{\nu}=\prod_{i=1}^{4}\xi(\mathbf{\Gamma_{i}}),
\end{eqnarray}
where $\xi(\mathbf{\Gamma_{i}})$ denotes the parity eigenvalue of the two lower Kramers degenerate bands at
the TRI momenta $\mathbf{\Gamma_{i}}$ ($\mathbf{\Gamma_{i}}=-\mathbf{\Gamma_{i}}$ up to a reciprocal lattice vector).
For $H_{0}(\bk)$, it is readily found that away from the critical lines $m_{0}\pm m_{1}\pm m_{2}=0$ where
the bulk gap gets closed, we have
\begin{eqnarray}
(-1)^{\nu}=\prod_{\alpha=\pm1, \beta=\pm1}\text{sgn}(m+\alpha m_{1}+\beta m_{2}).
\end{eqnarray}
According to the above formula,  the phase diagram can be straightforwardly determined,
as shown in Fig.\ref{phase}(a). In the phase diagram, the normal (or trivial) insulator (NI)
phase is further divided into two distinct parts. The first part labeled as NI$_{1}$ has zero (mod $4$) band
inversion (or equivalently to say, the parity eigenvalues at the four TRI momenta take the same sign in these regimes),
and the second part labeled as NI$_{2}$ has band inversions at two TRI momenta.
We are interested in NI$_{2}$ since gapless boundary states can appear on certain boundaries, whereas they
are completely absent in NI$_{1}$.

In the following, we consider a specific case where the band inversion
occurs at the two TRI momenta $(k_{1},k_{2})=(0,0)$ and $(\pi,0)$ to illustrate the essential physics.
Without loss of generality, we take $m_{1}=1$ as the energy unit and set $m_{0}=0$ and $m_{2}=2$
to realize the desired condition. To reveal the selective existence of helical gapless states on certain boundaries,
we consider that the insulator takes a cylinder geometry with open boundary condition in one direction and
periodic boundary condition in the other orthogonal direction. The results shown in Figs.\ref{phase}(b)(c)
indicate that helical gapless states do not appear on the $x_{1}$-normal edges, but appear on the $x_{2}$-normal edges.
One can see that the helical edge states cross at both TRI momenta in the edge Brillouin zone, suggesting
the presence of two pairs of helical gapless states on each $x_{2}$-normal edge. However, unlike TIs,
the helical gapless edge states do not traverse the bulk gap, instead they form two floating bands within the gap.
It is noteworthy that conduction-valence asymmetry only affects the dispersion of
the floating bands. The floating bands exist even if the asymmetry is strong enough to change
the insulator to a metal, as shown in Fig.\ref{phase}(d).

Let us now take into account the $s_{\pm}$-wave superconductivity.
Before focusing on the boundary, we first discuss the bulk.
As the TRI BdG Hamiltonian belongs to the DIII class,
the band topology of its 2D bulk also follows a $Z_{2}$ classification\cite{schnyder2008,kitaev2009,Haim2019review}.
For the concerned spin-singlet pairing,  the $Z_{2}$ invariant is simply given by\cite{qi2010d}
\begin{eqnarray}
N_{\rm 2D}=\prod_{s}[\text{sgn}(\Delta_{s})]^{m_{s}},\label{invariant2}
\end{eqnarray}
where $m_{s}$ denotes the number of TRI momenta enclosed by the $s$th Fermi surface,
and $\text{sgn}(\Delta_{s})$ denotes the sign of the pairing on the $s$th Fermi surface.
Because the normal state has both time-reversal symmetry and
inversion symmetry, the Kramers degeneracy at every momentum forces the double degeneracy of Fermi surface, if any.
As a result, $N_{\rm 2D}$ is necessitated to take the trivial value $1$. Therefore, the bulk
is always topologically trivial for the concerned spin-singlet pairing. However, the inversion symmetry
is broken on the boundary, which consequently lifts the Kramers degeneracy.
Indeed, on each $x_{2}$-normal edge, the floating bands shown in Figs.\ref{phase}(c)(d)
do not have Kramers degeneracy away from the two TRI momenta, which, as will be shown in
the following,
enables the realization of 1D TRI TSC on the boundary.

\begin{figure}[t!]
\subfigure{\includegraphics[width=9cm, height=4cm]{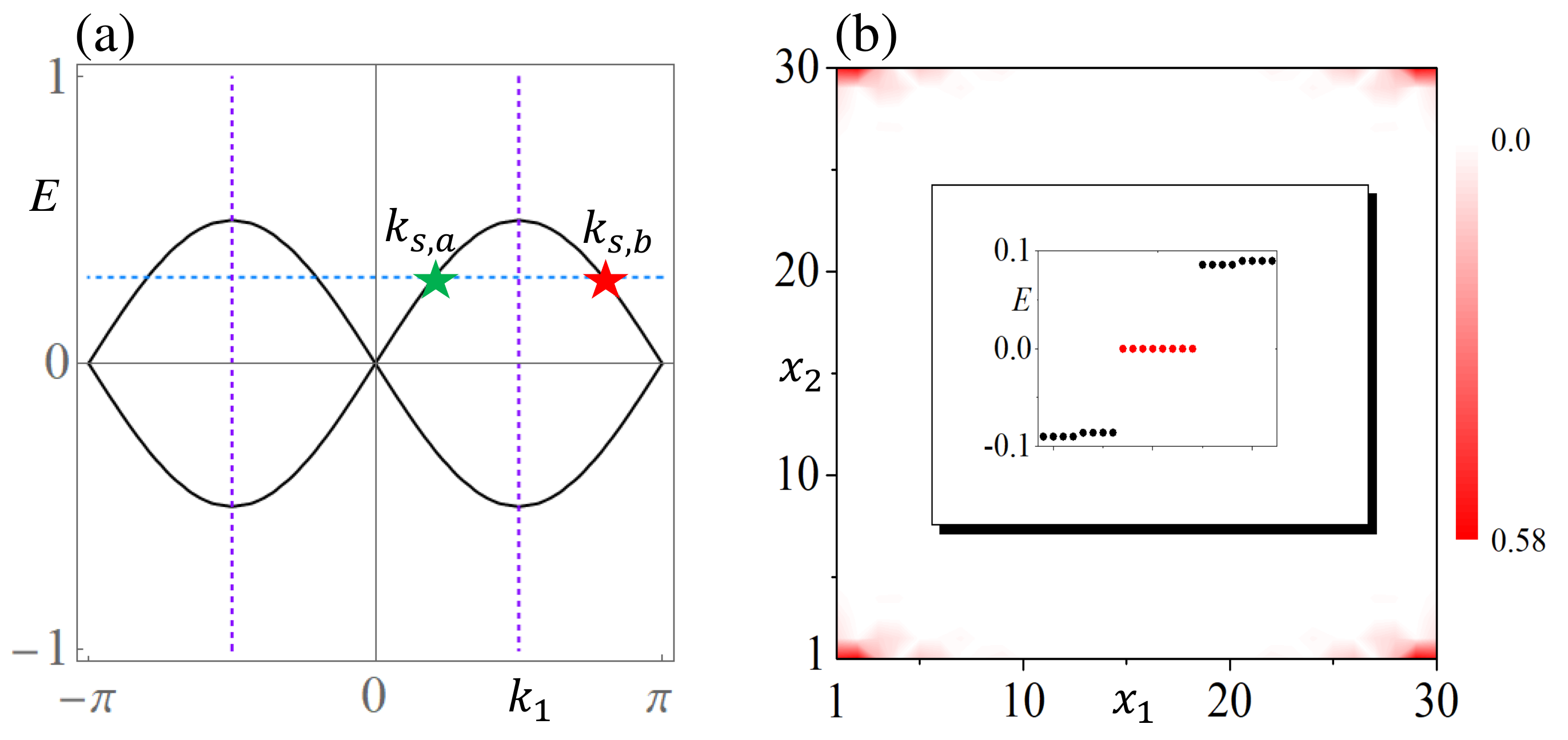}}
\caption{ (a) The sign of pairing on the Fermi points between $0$ and $\pi$.
The two solid black lines denote the floating bands formed by the gapless boundary states.
The dashed blue line denotes the Fermi level, and the two dashed purple lines indicate
the momenta at which the pairing changes sign. The blue and red star denote the Fermi points
with negative and positive pairing, respectively. (b) Density profiles of four Majorana Kramers pairs
are located at the four corners of the considered square sample. The inset shows a few energy eigenvalues closest
to zero. The parameters in (a) and (b) are $m_{0}=0$, $m_{1}=1$, $m_{2}=2$, $\lambda_{1}=0.5$,
$\lambda_{2}=1.0$, $\epsilon_{1,2}=0$, $\Delta_{1}=0$, $\Delta_{1}=0.2$.
}  \label{bt1}
\end{figure}

As the floating bands extends over the whole Brillouin zone, they can be described by a truly 1D lattice Hamiltonian.
This is sharply distinct to 2D TIs in which a lattice realization of the helical gapless edge states is known to be impossible.
Without loss of generality, we take the simple conduction-valence symmetric case for illustration.
In this limit, the floating bands shown in Fig.\ref{phase}(c) on one of the $x_{2}$-normal edges
is simply described by
\begin{eqnarray}
H_{f}(k_{1})=\lambda_{1}\sin k_{1}s_{3}.\label{ef}
\end{eqnarray}
In the presence of $s_{\pm}$-wave superconductivity, the
BdG Hamiltonian on the corresponding  edge reads
\begin{eqnarray}
H_{\rm e}(k_{1})&=&\lambda_{1}\sin k_{1}\tau_{0}\otimes s_{3}-\mu\tau_{3}\otimes s_{0}\nonumber\\
&&+(\Delta_{0}-\Delta_{1}\cos k_{1}-\Delta_{2}(k_{1}))\tau_{2}\otimes s_{2},\label{BH}
\end{eqnarray}
where the Pauli matrices $\tau_{i}$ act on the particle-hole space, and $\Delta_{2}(k_{1})\simeq\Delta_{1}(m_{0}-m_{1}\cos k_{1})/m_{2}$ (here
we provide a general expression),
which is originated from the $\Delta_{1}\cos k_{2}$ term of the pairing under the open boundary condition
in the $x_{2}$ direction (see details in the Supplemental Material\cite{supplemental}). This 1D TRI BdG Hamiltonian
also follows a $Z_{2}$ classification and the $Z_{2}$ invariant takes a form similar to Eq.(\ref{invariant2})\cite{qi2010d},
\begin{eqnarray}
N_{\rm 1D}=\prod_{s}[\text{sgn}(\Delta_{s})],\label{invariant1}
\end{eqnarray}
where $\text{sgn}(\Delta_{s})$ denotes the sign of the pairing on the $s$th Fermi point between
$0$ and $\pi$, as illustrated in Fig.\ref{bt1}(a). According to Eq.(\ref{BH}), there are two Fermi points
between $0$ and $\pi$ when $\mu\in(-\lambda_{1},\lambda_{1})$, which are located at
$k_{s,a}=\arcsin(|\mu/\lambda_{1}|)$ and $k_{s,b}=\pi-k_{s,a}$.
Following Eq.(\ref{invariant1}), we then have
\begin{eqnarray}
N_{\rm 1D}=\text{sgn}[(\Delta_{0}-\frac{m_{0}\Delta_{1}}{m_{2}})^{2}
-\Delta_{1}^{2}(1-\frac{m_{1}}{m_{2}})^{2}(1-\frac{\mu^{2}}{\lambda_{1}^{2}})^{2}]\,\label{invaraint3}
\end{eqnarray}
in the regime $\mu\in(-\lambda_{1},\lambda_{1})$. Under appropriate condition, $N_{\rm 1D}$
can take the nontrivial value $-1$, which corresponds to the realization of
a 1D BTRITSC. It is noteworthy that when $\mu^{2}>\lambda_{1}^{2}$,
there is no Fermi point, so $N_{\rm 1D}$ always takes the trivial value $1$,
indicating a trivial boundary.

To further demonstrate the above analytical results, we consider $\{m_{0},m_{1},m_{2},\Delta_{0},\Delta_{1}\}=\{0,1,2,0,0.2\}$.
 Then according to Eq.(\ref{invaraint3}), we have $N_{\rm 1D}=-1$
in the regime $\mu\in(-\lambda_{1},\lambda_{1})$. As a 1D TRI TSC is characterized
by the existence of one Majorana Kramers pair on each end\cite{Wong2012majorana,zhang2013kramers,keselman2013,haim2014,Gaidamauskas2014,Schrade2015}, the realization of BTRITSCs will be
manifested by the presence of Majorana Kramers pairs at the boundary of the BTRITSCs, i.e., the corners
of a square sample. As shown in Fig.\ref{bt1}(b), the numerical result confirms the prediction.
It is noteworthy that from a bulk perspective, the presence of Majorana Kramers pairs at the corners indicates
that the whole system is a second-order TRI TSC\cite{Yan2018hosc,Wang2018hosc}.

\begin{figure}
\subfigure{\includegraphics[width=9cm, height=9cm]{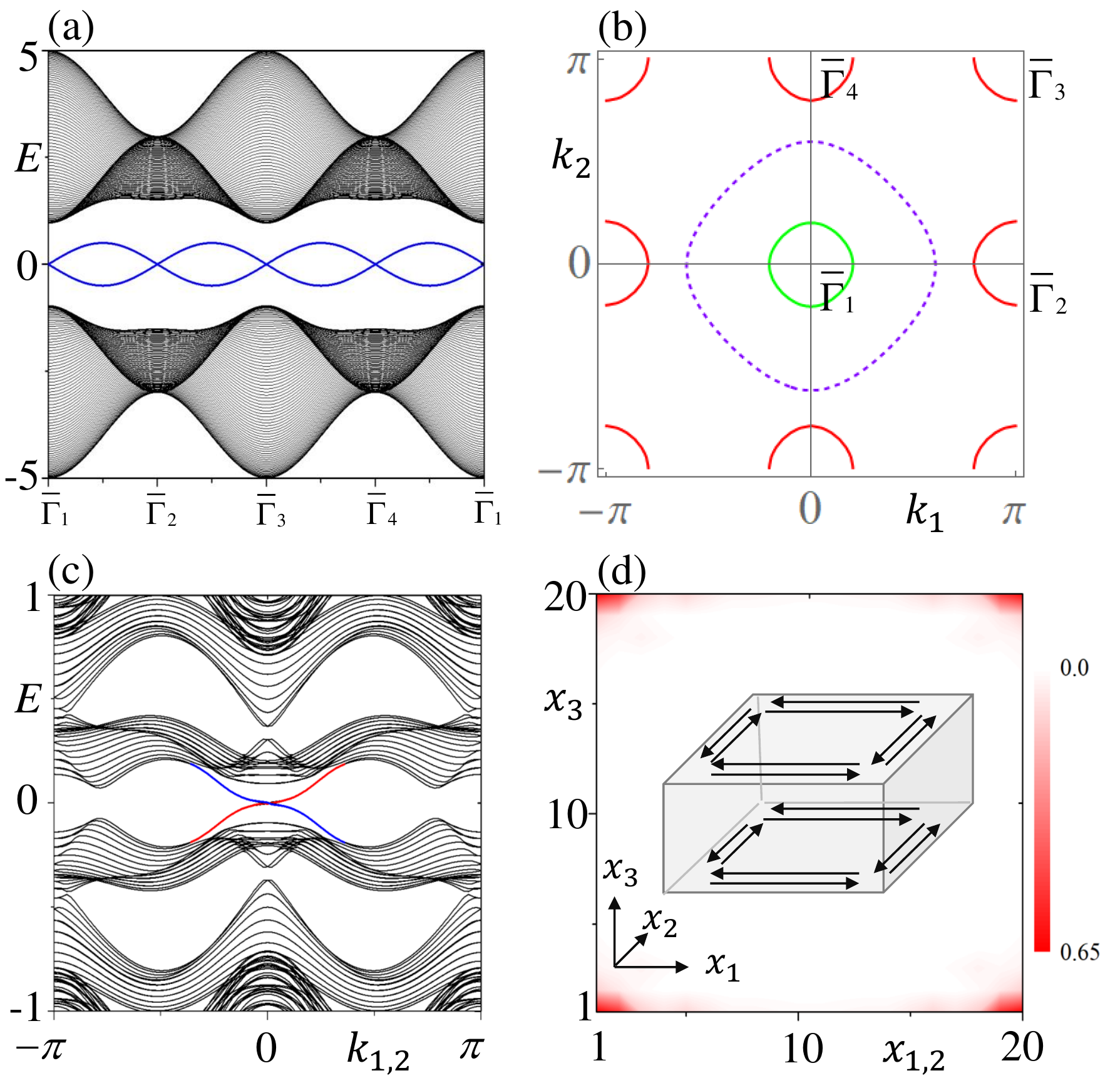}}
\caption{ (a) Normal-state energy spectrum for a geometry with open boundary condition in the
$x_{3}$ direction and periodic boundary condition in the $x_{1}$ and $x_{2}$ directions. The spectrum
is shown along the high symmetric lines of the surface Brillouin zone. (b) A configuration
with $N_{\rm 2D}=-1$. The dashed purple line denotes
the pairing nodal line, and
the solid red and green lines on its two sides denote Fermi surfaces with positive and negative pairing,
respectively.  (c) Superconducting-state energy spectrum for the configuration in (b).
The sample takes open boundary condition
in the $x_{1}(x_{2})$ and $x_{3}$ directions and periodic boundary condition in the $x_{2}(x_{1})$ direction.
Because the Hamiltonian has $C_{4}$ rotation symmetry, the energy spectra for the two cases
are the same. The in-gap dispersions are of four-fold degeneracy, which correspond to four pairs of
helical Majorana modes.
(d) The density profiles of the helical Majorana modes are localized at the hinges
of the sample. The inset provides an intuitive illustration of their distribution on the cubic
sample. Common parameters
are $\epsilon_{1,2,3}=0$, $m_{0}=0$, $m_{1,2}=1$, $m_{3}=3$, $\lambda_{1,2}=0.5$, $\lambda_{3}=1.5$.
In (b)(c)(d), $\Delta_{0}=0.1$, $\Delta_{1}=0.15$ and $\mu=0.3$.
}  \label{bt2}
\end{figure}

{\it BTRITSCs in 3D.---} The generalization to 3D is straightforward. We only need to generalize
the normal-state Hamiltonian into a 3D form and keep the pairing term intact.
Here we consider
\begin{eqnarray}
H_{0}(\bk)&=&\epsilon_{\rm 3D}(\bk)\Sigma_{00}+m_{\rm 3D}(\bk)\Sigma_{03}
+\sum_{i}\lambda_{i}\sin k_{i}\Sigma_{i1},
\end{eqnarray}
where $\epsilon_{\rm 3D}(\bk)=\sum_{i}\epsilon_{i}\cos k_{i}$ and
$m_{\rm 3D}(\bk)=m_{0}-\sum_{i}m_{i}\cos k_{i}$ ($i$ runs over
$1$, $2$ and $3$). Similarly, without loss of generality, we consider that the band inversion occurs
at the four TRI momenta of the $k_{3}=0$ plane. For such a configuration, both the strong
and weak $Z_{2}$ indices are trivial because the product of parity eigenvalues
in each of the $k_{1,2,3}=0/\pi$ planes gives the trivial value\cite{fu2007a}.
To realize this configuration,
we take $m_{0}=0$, $m_{1}=m_{2}=1$ and $m_{3}=3$. As shown in Fig.\ref{bt2}(a),
this configuration realizes 2D spin-degeneracy-lifted floating bands
on the $x_{3}$-normal surfaces. By performing similar analysis as in 2D, we find that the floating bands of the normal state
are described by
\begin{eqnarray}
H_{f}(k_{1},k_{2})=\lambda_{1}\sin k_{1}s_{1}+\lambda_{2}\sin k_{2}s_{2}.
\end{eqnarray}
In the presence of $s_{\pm}$-wave superconductivity, the corresponding surface BdG Hamiltonian takes a very simple form,
which reads
\begin{eqnarray}
H_{\rm s}(k_{1},k_{2})&=&\lambda_{1}\sin k_{1}\tau_{0}\otimes s_{1}+\lambda_{2}\sin k_{2}\tau_{3}\otimes s_{2}-\mu\tau_{3}\otimes s_{0}\nonumber\\
&&+(\Delta_{0}-\Delta_{1}(\cos k_{1}+\cos k_{2}))\tau_{2}\otimes s_{2}.\label{SH}
\end{eqnarray}
The band topology of $H_{\rm s}$ is just characterized by the $Z_{2}$ invariant given in Eq.(\ref{invariant2}).
As the normal-state Fermi surface is determined by
$\pm\sqrt{\sum_{j=1,2}\lambda_{j}^{2}\sin^{2} k_{j}}=\mu$, and the pairing
changes sign at the nodal line determined by $\Delta_{0}-\Delta_{1}(\cos k_{1}+\cos k_{2})=0$,
$N_{\rm 2D}$ can be intuitively determined by inspecting the configuration of
Fermi surface and pairing nodal line, as illustrated in Fig.\ref{bt2}(b). When the pairing
nodal line encloses one Fermi surface, $N_{\rm 2D}=-1$, and a 2D BTRITSC is realized. Similarly,  the
realization of a 2D BTRITSC is manifested by the presence
of helical Majorana modes\cite{Deng2012TRITSC,zhang2013kramers}, as shown in
Figs.\ref{bt2}(c)(d).
As the helical Majorana modes appear at the boundary
of the $z$-normal surfaces, the whole system is a 3D second-order TRI TSC from a bulk perspective\cite{Zhang2019hinge}.

{\it Effect of an external magnetic field.---} Thus far,
the time-reversal symmetry, which prohibits two time-reversal partner Majorana modes from coupling,
has been assumed to be preserved. Applying a magnetic field will generate a Zeeman term
of the form $(B_{1}\tau_{0}\otimes s_{1}+B_{2}\tau_{3}\otimes s_{2}+B_{3}\tau_{0}\otimes s_{3})\otimes \sigma_{0}$, accordingly breaking
this symmetry. Although the Majorana Kramers pairs in 2D and helical Majorana hinge modes in 3D
are no longer protected when the time-reversal symmetry is broken,
the magnetic field can induce interesting topological phase transitions.
In 2D, the Majorana Kramers pairs can be changed to solitary MZMs when the magnetic field exceeds
a critical value\cite{Yan2018hosc}. Remarkably, in 3D, a magnetic field in the $x_{1}$-$x_{2}$ plane can immediately gap the helical Majorana hinge modes and lead to the presence of solitary
MZMs at certain inversion-related corners\cite{Zhu2018hosc,Volpez2019SOTSC}. It means
that the magnetic field can change the second-order TRI TSC to a third-order time-reversal-symmetry-breaking TSC\cite{Yan2019hosca,Ahn2020hosc}.
It is noteworthy that such a response to magnetic field
is sharply distinct to second-order TRI TSCs realized by a combination of strong TIs and $s_{\pm}$-wave superconductivity\cite{Zhang2019hinge,kheirkhah2020vortex}.
For the latter, counter-intuitively, the magnetic field  cannot gap  the helical Majorana hinge modes, because they have a domain-wall origin therein
and the magnetic field cannot directly act on the domain-wall subspace\cite{kheirkhah2020vortex}.

{\it Conclusions.---} In this work, we have shown that
1D and 2D TRI
TSCs can be respectively realized on the boundary of 2D and 3D trivial band-inverted
insulators when their unprotected gapless boundary states are gapped by $s_{\pm}$-wave superconductivity.
Because the dimension of the BTRITSCs is lower than
the bulk by one, the BTRITSCs open a new route for the realization
of second-order TRI TSCs. In addition, we found that by applying a magnetic field,
a third-order TSC with Majorana corner modes can be readily induced from the second-order
TRI TSC realized in this route. Our established new scenario unveils that the widely-overlooked
trivial band-inverted insulators can also be applied for the realization of TSCs and concomitant Majorana modes,
hopefully broadening the scope of material candidates for TSCs.

{\it Acknowledgements.---} This work is supported by the Startup Grant (No. 74130-
18841219) and the National Science Foundation of
China (Grant No. 11904417).

\bibliography{dirac}

\begin{thebibliography}{85}%
\makeatletter
\providecommand \@ifxundefined [1]{%
 \@ifx{#1\undefined}
}%
\providecommand \@ifnum [1]{%
 \ifnum #1\expandafter \@firstoftwo
 \else \expandafter \@secondoftwo
 \fi
}%
\providecommand \@ifx [1]{%
 \ifx #1\expandafter \@firstoftwo
 \else \expandafter \@secondoftwo
 \fi
}%
\providecommand \natexlab [1]{#1}%
\providecommand \enquote  [1]{``#1''}%
\providecommand \bibnamefont  [1]{#1}%
\providecommand \bibfnamefont [1]{#1}%
\providecommand \citenamefont [1]{#1}%
\providecommand \href@noop [0]{\@secondoftwo}%
\providecommand \href [0]{\begingroup \@sanitize@url \@href}%
\providecommand \@href[1]{\@@startlink{#1}\@@href}%
\providecommand \@@href[1]{\endgroup#1\@@endlink}%
\providecommand \@sanitize@url [0]{\catcode `\\12\catcode `\$12\catcode
  `\&12\catcode `\#12\catcode `\^12\catcode `\_12\catcode `\%12\relax}%
\providecommand \@@startlink[1]{}%
\providecommand \@@endlink[0]{}%
\providecommand \url  [0]{\begingroup\@sanitize@url \@url }%
\providecommand \@url [1]{\endgroup\@href {#1}{\urlprefix }}%
\providecommand \urlprefix  [0]{URL }%
\providecommand \Eprint [0]{\href }%
\providecommand \doibase [0]{http://dx.doi.org/}%
\providecommand \selectlanguage [0]{\@gobble}%
\providecommand \bibinfo  [0]{\@secondoftwo}%
\providecommand \bibfield  [0]{\@secondoftwo}%
\providecommand \translation [1]{[#1]}%
\providecommand \BibitemOpen [0]{}%
\providecommand \bibitemStop [0]{}%
\providecommand \bibitemNoStop [0]{.\EOS\space}%
\providecommand \EOS [0]{\spacefactor3000\relax}%
\providecommand \BibitemShut  [1]{\csname bibitem#1\endcsname}%
\let\auto@bib@innerbib\@empty
\bibitem [{\citenamefont {Hasan}\ and\ \citenamefont {Kane}(2010)}]{hasan2010}%
  \BibitemOpen
  \bibfield  {author} {\bibinfo {author} {\bibfnamefont {M.~Z.}\ \bibnamefont
  {Hasan}}\ and\ \bibinfo {author} {\bibfnamefont {C.~L.}\ \bibnamefont
  {Kane}},\ }\bibfield  {title} {\enquote {\bibinfo {title}
  {\textit{Colloquium} : Topological insulators},}\ }\href {\doibase
  10.1103/RevModPhys.82.3045} {\bibfield  {journal} {\bibinfo  {journal} {Rev.
  Mod. Phys.}\ }\textbf {\bibinfo {volume} {82}},\ \bibinfo {pages}
  {3045--3067} (\bibinfo {year} {2010})}\BibitemShut {NoStop}%
\bibitem [{\citenamefont {Qi}\ and\ \citenamefont {Zhang}(2011)}]{qi2011}%
  \BibitemOpen
  \bibfield  {author} {\bibinfo {author} {\bibfnamefont {Xiao-Liang}\
  \bibnamefont {Qi}}\ and\ \bibinfo {author} {\bibfnamefont {Shou-Cheng}\
  \bibnamefont {Zhang}},\ }\bibfield  {title} {\enquote {\bibinfo {title}
  {Topological insulators and superconductors},}\ }\href {\doibase
  10.1103/RevModPhys.83.1057} {\bibfield  {journal} {\bibinfo  {journal} {Rev.
  Mod. Phys.}\ }\textbf {\bibinfo {volume} {83}},\ \bibinfo {pages}
  {1057--1110} (\bibinfo {year} {2011})}\BibitemShut {NoStop}%
\bibitem [{\citenamefont {Kane}\ and\ \citenamefont {Mele}(2005)}]{Kane2005b}%
  \BibitemOpen
  \bibfield  {author} {\bibinfo {author} {\bibfnamefont {C.~L.}\ \bibnamefont
  {Kane}}\ and\ \bibinfo {author} {\bibfnamefont {E.~J.}\ \bibnamefont
  {Mele}},\ }\bibfield  {title} {\enquote {\bibinfo {title} {${Z}_{2}$
  topological order and the quantum spin hall effect},}\ }\href {\doibase
  10.1103/PhysRevLett.95.146802} {\bibfield  {journal} {\bibinfo  {journal}
  {Phys. Rev. Lett.}\ }\textbf {\bibinfo {volume} {95}},\ \bibinfo {pages}
  {146802} (\bibinfo {year} {2005})}\BibitemShut {NoStop}%
\bibitem [{\citenamefont {Fu}\ \emph {et~al.}(2007)\citenamefont {Fu},
  \citenamefont {Kane},\ and\ \citenamefont {Mele}}]{fu2007b}%
  \BibitemOpen
  \bibfield  {author} {\bibinfo {author} {\bibfnamefont {Liang}\ \bibnamefont
  {Fu}}, \bibinfo {author} {\bibfnamefont {C.~L.}\ \bibnamefont {Kane}}, \ and\
  \bibinfo {author} {\bibfnamefont {E.~J.}\ \bibnamefont {Mele}},\ }\bibfield
  {title} {\enquote {\bibinfo {title} {Topological insulators in three
  dimensions},}\ }\href {\doibase 10.1103/PhysRevLett.98.106803} {\bibfield
  {journal} {\bibinfo  {journal} {Phys. Rev. Lett.}\ }\textbf {\bibinfo
  {volume} {98}},\ \bibinfo {eid} {106803} (\bibinfo {year}
  {2007})}\BibitemShut {NoStop}%
\bibitem [{\citenamefont {Moore}\ and\ \citenamefont
  {Balents}(2007)}]{moore2007}%
  \BibitemOpen
  \bibfield  {author} {\bibinfo {author} {\bibfnamefont {J.~E.}\ \bibnamefont
  {Moore}}\ and\ \bibinfo {author} {\bibfnamefont {L.}~\bibnamefont
  {Balents}},\ }\bibfield  {title} {\enquote {\bibinfo {title} {Topological
  invariants of time-reversal-invariant band structures},}\ }\href {\doibase
  10.1103/PhysRevB.75.121306} {\bibfield  {journal} {\bibinfo  {journal} {Phys.
  Rev. B}\ }\textbf {\bibinfo {volume} {75}},\ \bibinfo {eid} {121306}
  (\bibinfo {year} {2007})}\BibitemShut {NoStop}%
\bibitem [{\citenamefont {Roy}(2009)}]{roy2009}%
  \BibitemOpen
  \bibfield  {author} {\bibinfo {author} {\bibfnamefont {Rahul}\ \bibnamefont
  {Roy}},\ }\bibfield  {title} {\enquote {\bibinfo {title} {Topological phases
  and the quantum spin hall effect in three dimensions},}\ }\href {\doibase
  10.1103/PhysRevB.79.195322} {\bibfield  {journal} {\bibinfo  {journal} {Phys.
  Rev. B}\ }\textbf {\bibinfo {volume} {79}},\ \bibinfo {pages} {195322}
  (\bibinfo {year} {2009})}\BibitemShut {NoStop}%
\bibitem [{\citenamefont {Fu}\ and\ \citenamefont {Kane}(2007)}]{fu2007a}%
  \BibitemOpen
  \bibfield  {author} {\bibinfo {author} {\bibfnamefont {Liang}\ \bibnamefont
  {Fu}}\ and\ \bibinfo {author} {\bibfnamefont {C.~L.}\ \bibnamefont {Kane}},\
  }\bibfield  {title} {\enquote {\bibinfo {title} {Topological insulators with
  inversion symmetry},}\ }\href {\doibase 10.1103/PhysRevB.76.045302}
  {\bibfield  {journal} {\bibinfo  {journal} {PRB}\ }\textbf {\bibinfo {volume}
  {76}},\ \bibinfo {eid} {045302} (\bibinfo {year} {2007})}\BibitemShut
  {NoStop}%
\bibitem [{\citenamefont {K{\"o}nig}\ \emph {et~al.}(2007)\citenamefont
  {K{\"o}nig}, \citenamefont {Wiedmann}, \citenamefont {Br{\"u}ne},
  \citenamefont {Roth}, \citenamefont {Buhmann}, \citenamefont {Molenkamp},
  \citenamefont {Qi},\ and\ \citenamefont {Zhang}}]{Konig2007}%
  \BibitemOpen
  \bibfield  {author} {\bibinfo {author} {\bibfnamefont {Markus}\ \bibnamefont
  {K{\"o}nig}}, \bibinfo {author} {\bibfnamefont {Steffen}\ \bibnamefont
  {Wiedmann}}, \bibinfo {author} {\bibfnamefont {Christoph}\ \bibnamefont
  {Br{\"u}ne}}, \bibinfo {author} {\bibfnamefont {Andreas}\ \bibnamefont
  {Roth}}, \bibinfo {author} {\bibfnamefont {Hartmut}\ \bibnamefont {Buhmann}},
  \bibinfo {author} {\bibfnamefont {Laurens~W.}\ \bibnamefont {Molenkamp}},
  \bibinfo {author} {\bibfnamefont {Xiao-Liang}\ \bibnamefont {Qi}}, \ and\
  \bibinfo {author} {\bibfnamefont {Shou-Cheng}\ \bibnamefont {Zhang}},\
  }\bibfield  {title} {\enquote {\bibinfo {title} {Quantum spin hall insulator
  state in hgte quantum wells},}\ }\href {\doibase 10.1126/science.1148047}
  {\bibfield  {journal} {\bibinfo  {journal} {Science}\ }\textbf {\bibinfo
  {volume} {318}},\ \bibinfo {pages} {766--770} (\bibinfo {year} {2007})},\
  \Eprint
  {http://arxiv.org/abs/https://science.sciencemag.org/content/318/5851/766.full.pdf}
  {https://science.sciencemag.org/content/318/5851/766.full.pdf} \BibitemShut
  {NoStop}%
\bibitem [{\citenamefont {Knez}\ \emph {et~al.}(2011)\citenamefont {Knez},
  \citenamefont {Du},\ and\ \citenamefont {Sullivan}}]{Knez2011}%
  \BibitemOpen
  \bibfield  {author} {\bibinfo {author} {\bibfnamefont {Ivan}\ \bibnamefont
  {Knez}}, \bibinfo {author} {\bibfnamefont {Rui-Rui}\ \bibnamefont {Du}}, \
  and\ \bibinfo {author} {\bibfnamefont {Gerard}\ \bibnamefont {Sullivan}},\
  }\bibfield  {title} {\enquote {\bibinfo {title} {Evidence for helical edge
  modes in inverted $\mathrm{InAs}/\mathrm{GaSb}$ quantum wells},}\ }\href
  {\doibase 10.1103/PhysRevLett.107.136603} {\bibfield  {journal} {\bibinfo
  {journal} {Phys. Rev. Lett.}\ }\textbf {\bibinfo {volume} {107}},\ \bibinfo
  {pages} {136603} (\bibinfo {year} {2011})}\BibitemShut {NoStop}%
\bibitem [{\citenamefont {Wu}\ \emph {et~al.}(2018)\citenamefont {Wu},
  \citenamefont {Fatemi}, \citenamefont {Gibson}, \citenamefont {Watanabe},
  \citenamefont {Taniguchi}, \citenamefont {Cava},\ and\ \citenamefont
  {Jarillo-Herrero}}]{Wu2018QSHE}%
  \BibitemOpen
  \bibfield  {author} {\bibinfo {author} {\bibfnamefont {Sanfeng}\ \bibnamefont
  {Wu}}, \bibinfo {author} {\bibfnamefont {Valla}\ \bibnamefont {Fatemi}},
  \bibinfo {author} {\bibfnamefont {Quinn~D.}\ \bibnamefont {Gibson}}, \bibinfo
  {author} {\bibfnamefont {Kenji}\ \bibnamefont {Watanabe}}, \bibinfo {author}
  {\bibfnamefont {Takashi}\ \bibnamefont {Taniguchi}}, \bibinfo {author}
  {\bibfnamefont {Robert~J.}\ \bibnamefont {Cava}}, \ and\ \bibinfo {author}
  {\bibfnamefont {Pablo}\ \bibnamefont {Jarillo-Herrero}},\ }\bibfield  {title}
  {\enquote {\bibinfo {title} {Observation of the quantum spin hall effect up
  to 100 kelvin in a monolayer crystal},}\ }\href {\doibase
  10.1126/science.aan6003} {\bibfield  {journal} {\bibinfo  {journal}
  {Science}\ }\textbf {\bibinfo {volume} {359}},\ \bibinfo {pages} {76--79}
  (\bibinfo {year} {2018})}\BibitemShut {NoStop}%
\bibitem [{\citenamefont {Zhang}\ \emph {et~al.}(2009)\citenamefont {Zhang},
  \citenamefont {Liu}, \citenamefont {Qi}, \citenamefont {Dai}, \citenamefont
  {Fang},\ and\ \citenamefont {Zhang}}]{Zhang2009bise}%
  \BibitemOpen
  \bibfield  {author} {\bibinfo {author} {\bibfnamefont {Haijun}\ \bibnamefont
  {Zhang}}, \bibinfo {author} {\bibfnamefont {Chao-Xing}\ \bibnamefont {Liu}},
  \bibinfo {author} {\bibfnamefont {Xiao-Liang}\ \bibnamefont {Qi}}, \bibinfo
  {author} {\bibfnamefont {Xi}~\bibnamefont {Dai}}, \bibinfo {author}
  {\bibfnamefont {Zhong}\ \bibnamefont {Fang}}, \ and\ \bibinfo {author}
  {\bibfnamefont {Shou-Cheng}\ \bibnamefont {Zhang}},\ }\bibfield  {title}
  {\enquote {\bibinfo {title} {Topological insulators in bi2se3, bi2te3 and
  sb2te3 with a single dirac cone on the surface},}\ }\href {\doibase
  10.1038/nphys1270} {\bibfield  {journal} {\bibinfo  {journal} {Nature
  Physics}\ }\textbf {\bibinfo {volume} {5}},\ \bibinfo {pages} {438--442}
  (\bibinfo {year} {2009})}\BibitemShut {NoStop}%
\bibitem [{\citenamefont {Chen}\ \emph {et~al.}(2009)\citenamefont {Chen},
  \citenamefont {Analytis}, \citenamefont {Chu}, \citenamefont {Liu},
  \citenamefont {Mo}, \citenamefont {Qi}, \citenamefont {Zhang}, \citenamefont
  {Lu}, \citenamefont {Dai}, \citenamefont {Fang}, \citenamefont {Zhang},
  \citenamefont {Fisher}, \citenamefont {Hussain},\ and\ \citenamefont
  {Shen}}]{Chen2009bise}%
  \BibitemOpen
  \bibfield  {author} {\bibinfo {author} {\bibfnamefont {Y.~L.}\ \bibnamefont
  {Chen}}, \bibinfo {author} {\bibfnamefont {J.~G.}\ \bibnamefont {Analytis}},
  \bibinfo {author} {\bibfnamefont {J.-H.}\ \bibnamefont {Chu}}, \bibinfo
  {author} {\bibfnamefont {Z.~K.}\ \bibnamefont {Liu}}, \bibinfo {author}
  {\bibfnamefont {S.-K.}\ \bibnamefont {Mo}}, \bibinfo {author} {\bibfnamefont
  {X.~L.}\ \bibnamefont {Qi}}, \bibinfo {author} {\bibfnamefont {H.~J.}\
  \bibnamefont {Zhang}}, \bibinfo {author} {\bibfnamefont {D.~H.}\ \bibnamefont
  {Lu}}, \bibinfo {author} {\bibfnamefont {X.}~\bibnamefont {Dai}}, \bibinfo
  {author} {\bibfnamefont {Z.}~\bibnamefont {Fang}}, \bibinfo {author}
  {\bibfnamefont {S.~C.}\ \bibnamefont {Zhang}}, \bibinfo {author}
  {\bibfnamefont {I.~R.}\ \bibnamefont {Fisher}}, \bibinfo {author}
  {\bibfnamefont {Z.}~\bibnamefont {Hussain}}, \ and\ \bibinfo {author}
  {\bibfnamefont {Z.-X.}\ \bibnamefont {Shen}},\ }\bibfield  {title} {\enquote
  {\bibinfo {title} {Experimental realization of a three-dimensional
  topological insulator, bi2te3},}\ }\href {\doibase 10.1126/science.1173034}
  {\bibfield  {journal} {\bibinfo  {journal} {Science}\ }\textbf {\bibinfo
  {volume} {325}},\ \bibinfo {pages} {178--181} (\bibinfo {year} {2009})},\
  \Eprint
  {http://arxiv.org/abs/https://science.sciencemag.org/content/325/5937/178.full.pdf}
  {https://science.sciencemag.org/content/325/5937/178.full.pdf} \BibitemShut
  {NoStop}%
\bibitem [{\citenamefont {Xia}\ \emph {et~al.}(2009)\citenamefont {Xia},
  \citenamefont {Qian}, \citenamefont {Hsieh}, \citenamefont {Wray},
  \citenamefont {Pal}, \citenamefont {Lin}, \citenamefont {Bansil},
  \citenamefont {Grauer}, \citenamefont {Hor}, \citenamefont {Cava},\ and\
  \citenamefont {Hasan}}]{Xia2009bise}%
  \BibitemOpen
  \bibfield  {author} {\bibinfo {author} {\bibfnamefont {Y.}~\bibnamefont
  {Xia}}, \bibinfo {author} {\bibfnamefont {D.}~\bibnamefont {Qian}}, \bibinfo
  {author} {\bibfnamefont {D.}~\bibnamefont {Hsieh}}, \bibinfo {author}
  {\bibfnamefont {L.}~\bibnamefont {Wray}}, \bibinfo {author} {\bibfnamefont
  {A.}~\bibnamefont {Pal}}, \bibinfo {author} {\bibfnamefont {H.}~\bibnamefont
  {Lin}}, \bibinfo {author} {\bibfnamefont {A.}~\bibnamefont {Bansil}},
  \bibinfo {author} {\bibfnamefont {D.}~\bibnamefont {Grauer}}, \bibinfo
  {author} {\bibfnamefont {Y.~S.}\ \bibnamefont {Hor}}, \bibinfo {author}
  {\bibfnamefont {R.~J.}\ \bibnamefont {Cava}}, \ and\ \bibinfo {author}
  {\bibfnamefont {M.~Z.}\ \bibnamefont {Hasan}},\ }\bibfield  {title} {\enquote
  {\bibinfo {title} {Observation of a large-gap topological-insulator class
  with a single dirac cone on the surface},}\ }\href {\doibase
  10.1038/nphys1274} {\bibfield  {journal} {\bibinfo  {journal} {Nature
  Physics}\ }\textbf {\bibinfo {volume} {5}},\ \bibinfo {pages} {398--402}
  (\bibinfo {year} {2009})}\BibitemShut {NoStop}%
\bibitem [{\citenamefont {Hsieh}\ \emph
  {et~al.}(2009{\natexlab{a}})\citenamefont {Hsieh}, \citenamefont {Xia},
  \citenamefont {Qian}, \citenamefont {Wray}, \citenamefont {Meier},
  \citenamefont {Dil}, \citenamefont {Osterwalder}, \citenamefont {Patthey},
  \citenamefont {Fedorov}, \citenamefont {Lin}, \citenamefont {Bansil},
  \citenamefont {Grauer}, \citenamefont {Hor}, \citenamefont {Cava},\ and\
  \citenamefont {Hasan}}]{Hsieh2009bise}%
  \BibitemOpen
  \bibfield  {author} {\bibinfo {author} {\bibfnamefont {D.}~\bibnamefont
  {Hsieh}}, \bibinfo {author} {\bibfnamefont {Y.}~\bibnamefont {Xia}}, \bibinfo
  {author} {\bibfnamefont {D.}~\bibnamefont {Qian}}, \bibinfo {author}
  {\bibfnamefont {L.}~\bibnamefont {Wray}}, \bibinfo {author} {\bibfnamefont
  {F.}~\bibnamefont {Meier}}, \bibinfo {author} {\bibfnamefont {J.~H.}\
  \bibnamefont {Dil}}, \bibinfo {author} {\bibfnamefont {J.}~\bibnamefont
  {Osterwalder}}, \bibinfo {author} {\bibfnamefont {L.}~\bibnamefont
  {Patthey}}, \bibinfo {author} {\bibfnamefont {A.~V.}\ \bibnamefont
  {Fedorov}}, \bibinfo {author} {\bibfnamefont {H.}~\bibnamefont {Lin}},
  \bibinfo {author} {\bibfnamefont {A.}~\bibnamefont {Bansil}}, \bibinfo
  {author} {\bibfnamefont {D.}~\bibnamefont {Grauer}}, \bibinfo {author}
  {\bibfnamefont {Y.~S.}\ \bibnamefont {Hor}}, \bibinfo {author} {\bibfnamefont
  {R.~J.}\ \bibnamefont {Cava}}, \ and\ \bibinfo {author} {\bibfnamefont
  {M.~Z.}\ \bibnamefont {Hasan}},\ }\bibfield  {title} {\enquote {\bibinfo
  {title} {Observation of time-reversal-protected single-dirac-cone
  topological-insulator states in ${\mathrm{bi}}_{2}{\mathrm{te}}_{3}$ and
  ${\mathrm{sb}}_{2}{\mathrm{te}}_{3}$},}\ }\href {\doibase
  10.1103/PhysRevLett.103.146401} {\bibfield  {journal} {\bibinfo  {journal}
  {Phys. Rev. Lett.}\ }\textbf {\bibinfo {volume} {103}},\ \bibinfo {pages}
  {146401} (\bibinfo {year} {2009}{\natexlab{a}})}\BibitemShut {NoStop}%
\bibitem [{\citenamefont {Liu}\ \emph {et~al.}(2016)\citenamefont {Liu},
  \citenamefont {Zhou}, \citenamefont {Yao},\ and\ \citenamefont
  {Zhang}}]{Liu2016weak}%
  \BibitemOpen
  \bibfield  {author} {\bibinfo {author} {\bibfnamefont {Cheng-Cheng}\
  \bibnamefont {Liu}}, \bibinfo {author} {\bibfnamefont {Jin-Jian}\
  \bibnamefont {Zhou}}, \bibinfo {author} {\bibfnamefont {Yugui}\ \bibnamefont
  {Yao}}, \ and\ \bibinfo {author} {\bibfnamefont {Fan}\ \bibnamefont
  {Zhang}},\ }\bibfield  {title} {\enquote {\bibinfo {title} {Weak topological
  insulators and composite weyl semimetals:
  $\ensuremath{\beta}\text{\ensuremath{-}}{\mathrm{bi}}_{4}{X}_{4}$
  ($x=\mathrm{Br}$, i)},}\ }\href {\doibase 10.1103/PhysRevLett.116.066801}
  {\bibfield  {journal} {\bibinfo  {journal} {Phys. Rev. Lett.}\ }\textbf
  {\bibinfo {volume} {116}},\ \bibinfo {pages} {066801} (\bibinfo {year}
  {2016})}\BibitemShut {NoStop}%
\bibitem [{\citenamefont {Noguchi}\ \emph {et~al.}(2019)\citenamefont
  {Noguchi}, \citenamefont {Takahashi}, \citenamefont {Kuroda}, \citenamefont
  {Ochi}, \citenamefont {Shirasawa}, \citenamefont {Sakano}, \citenamefont
  {Bareille}, \citenamefont {Nakayama}, \citenamefont {Watson}, \citenamefont
  {Yaji}, \citenamefont {Harasawa}, \citenamefont {Iwasawa}, \citenamefont
  {Dudin}, \citenamefont {Kim}, \citenamefont {Hoesch}, \citenamefont
  {Kandyba}, \citenamefont {Giampietri}, \citenamefont {Barinov}, \citenamefont
  {Shin}, \citenamefont {Arita}, \citenamefont {Sasagawa},\ and\ \citenamefont
  {Kondo}}]{Noguchi2019}%
  \BibitemOpen
  \bibfield  {author} {\bibinfo {author} {\bibfnamefont {Ryo}\ \bibnamefont
  {Noguchi}}, \bibinfo {author} {\bibfnamefont {T.}~\bibnamefont {Takahashi}},
  \bibinfo {author} {\bibfnamefont {K.}~\bibnamefont {Kuroda}}, \bibinfo
  {author} {\bibfnamefont {M.}~\bibnamefont {Ochi}}, \bibinfo {author}
  {\bibfnamefont {T.}~\bibnamefont {Shirasawa}}, \bibinfo {author}
  {\bibfnamefont {M.}~\bibnamefont {Sakano}}, \bibinfo {author} {\bibfnamefont
  {C.}~\bibnamefont {Bareille}}, \bibinfo {author} {\bibfnamefont
  {M.}~\bibnamefont {Nakayama}}, \bibinfo {author} {\bibfnamefont {M.~D.}\
  \bibnamefont {Watson}}, \bibinfo {author} {\bibfnamefont {K.}~\bibnamefont
  {Yaji}}, \bibinfo {author} {\bibfnamefont {A.}~\bibnamefont {Harasawa}},
  \bibinfo {author} {\bibfnamefont {H.}~\bibnamefont {Iwasawa}}, \bibinfo
  {author} {\bibfnamefont {P.}~\bibnamefont {Dudin}}, \bibinfo {author}
  {\bibfnamefont {T.~K.}\ \bibnamefont {Kim}}, \bibinfo {author} {\bibfnamefont
  {M.}~\bibnamefont {Hoesch}}, \bibinfo {author} {\bibfnamefont
  {V.}~\bibnamefont {Kandyba}}, \bibinfo {author} {\bibfnamefont
  {A.}~\bibnamefont {Giampietri}}, \bibinfo {author} {\bibfnamefont
  {A.}~\bibnamefont {Barinov}}, \bibinfo {author} {\bibfnamefont
  {S.}~\bibnamefont {Shin}}, \bibinfo {author} {\bibfnamefont {R.}~\bibnamefont
  {Arita}}, \bibinfo {author} {\bibfnamefont {T.}~\bibnamefont {Sasagawa}}, \
  and\ \bibinfo {author} {\bibfnamefont {Takeshi}\ \bibnamefont {Kondo}},\
  }\bibfield  {title} {\enquote {\bibinfo {title} {A weak topological insulator
  state in quasi-one-dimensional bismuth iodide},}\ }\href {\doibase
  10.1038/s41586-019-0927-7} {\bibfield  {journal} {\bibinfo  {journal}
  {Nature}\ }\textbf {\bibinfo {volume} {566}},\ \bibinfo {pages} {518--522}
  (\bibinfo {year} {2019})}\BibitemShut {NoStop}%
\bibitem [{\citenamefont {Hsieh}\ \emph
  {et~al.}(2009{\natexlab{b}})\citenamefont {Hsieh}, \citenamefont {Xia},
  \citenamefont {Qian}, \citenamefont {Wray}, \citenamefont {Dil},
  \citenamefont {Meier}, \citenamefont {Osterwalder}, \citenamefont {Patthey},
  \citenamefont {Checkelsky}, \citenamefont {Ong}, \citenamefont {Fedorov},
  \citenamefont {Lin}, \citenamefont {Bansil}, \citenamefont {Grauer},
  \citenamefont {Hor}, \citenamefont {Cava},\ and\ \citenamefont
  {Hasan}}]{Hsieh2009helical}%
  \BibitemOpen
  \bibfield  {author} {\bibinfo {author} {\bibfnamefont {D.}~\bibnamefont
  {Hsieh}}, \bibinfo {author} {\bibfnamefont {Y.}~\bibnamefont {Xia}}, \bibinfo
  {author} {\bibfnamefont {D.}~\bibnamefont {Qian}}, \bibinfo {author}
  {\bibfnamefont {L.}~\bibnamefont {Wray}}, \bibinfo {author} {\bibfnamefont
  {J.~H.}\ \bibnamefont {Dil}}, \bibinfo {author} {\bibfnamefont
  {F.}~\bibnamefont {Meier}}, \bibinfo {author} {\bibfnamefont
  {J.}~\bibnamefont {Osterwalder}}, \bibinfo {author} {\bibfnamefont
  {L.}~\bibnamefont {Patthey}}, \bibinfo {author} {\bibfnamefont {J.~G.}\
  \bibnamefont {Checkelsky}}, \bibinfo {author} {\bibfnamefont {N.~P.}\
  \bibnamefont {Ong}}, \bibinfo {author} {\bibfnamefont {A.~V.}\ \bibnamefont
  {Fedorov}}, \bibinfo {author} {\bibfnamefont {H.}~\bibnamefont {Lin}},
  \bibinfo {author} {\bibfnamefont {A.}~\bibnamefont {Bansil}}, \bibinfo
  {author} {\bibfnamefont {D.}~\bibnamefont {Grauer}}, \bibinfo {author}
  {\bibfnamefont {Y.~S.}\ \bibnamefont {Hor}}, \bibinfo {author} {\bibfnamefont
  {R.~J.}\ \bibnamefont {Cava}}, \ and\ \bibinfo {author} {\bibfnamefont
  {M.~Z.}\ \bibnamefont {Hasan}},\ }\bibfield  {title} {\enquote {\bibinfo
  {title} {A tunable topological insulator in the spin helical dirac transport
  regime},}\ }\href {\doibase 10.1038/nature08234} {\bibfield  {journal}
  {\bibinfo  {journal} {Nature}\ }\textbf {\bibinfo {volume} {460}},\ \bibinfo
  {pages} {1101--1105} (\bibinfo {year} {2009}{\natexlab{b}})}\BibitemShut
  {NoStop}%
\bibitem [{\citenamefont {Fu}\ and\ \citenamefont {Kane}(2008)}]{fu2008}%
  \BibitemOpen
  \bibfield  {author} {\bibinfo {author} {\bibfnamefont {Liang}\ \bibnamefont
  {Fu}}\ and\ \bibinfo {author} {\bibfnamefont {C.~L.}\ \bibnamefont {Kane}},\
  }\bibfield  {title} {\enquote {\bibinfo {title} {Superconducting proximity
  effect and majorana fermions at the surface of a topological insulator},}\
  }\href {\doibase 10.1103/PhysRevLett.100.096407} {\bibfield  {journal}
  {\bibinfo  {journal} {Phys. Rev. Lett.}\ }\textbf {\bibinfo {volume} {100}},\
  \bibinfo {pages} {096407} (\bibinfo {year} {2008})}\BibitemShut {NoStop}%
\bibitem [{\citenamefont {Hosur}\ \emph {et~al.}(2011)\citenamefont {Hosur},
  \citenamefont {Ghaemi}, \citenamefont {Mong},\ and\ \citenamefont
  {Vishwanath}}]{Hosur2011MZM}%
  \BibitemOpen
  \bibfield  {author} {\bibinfo {author} {\bibfnamefont {Pavan}\ \bibnamefont
  {Hosur}}, \bibinfo {author} {\bibfnamefont {Pouyan}\ \bibnamefont {Ghaemi}},
  \bibinfo {author} {\bibfnamefont {Roger S.~K.}\ \bibnamefont {Mong}}, \ and\
  \bibinfo {author} {\bibfnamefont {Ashvin}\ \bibnamefont {Vishwanath}},\
  }\bibfield  {title} {\enquote {\bibinfo {title} {Majorana modes at the ends
  of superconductor vortices in doped topological insulators},}\ }\href
  {\doibase 10.1103/PhysRevLett.107.097001} {\bibfield  {journal} {\bibinfo
  {journal} {Phys. Rev. Lett.}\ }\textbf {\bibinfo {volume} {107}},\ \bibinfo
  {pages} {097001} (\bibinfo {year} {2011})}\BibitemShut {NoStop}%
\bibitem [{\citenamefont {Qin}\ \emph {et~al.}(2019)\citenamefont {Qin},
  \citenamefont {Hu}, \citenamefont {Wu}, \citenamefont {Dai}, \citenamefont
  {Fang}, \citenamefont {Zhang},\ and\ \citenamefont {Hu}}]{Qin2019vortex2}%
  \BibitemOpen
  \bibfield  {author} {\bibinfo {author} {\bibfnamefont {Shengshan}\
  \bibnamefont {Qin}}, \bibinfo {author} {\bibfnamefont {Lunhui}\ \bibnamefont
  {Hu}}, \bibinfo {author} {\bibfnamefont {Xianxin}\ \bibnamefont {Wu}},
  \bibinfo {author} {\bibfnamefont {Xia}\ \bibnamefont {Dai}}, \bibinfo
  {author} {\bibfnamefont {Chen}\ \bibnamefont {Fang}}, \bibinfo {author}
  {\bibfnamefont {Fu-Chun}\ \bibnamefont {Zhang}}, \ and\ \bibinfo {author}
  {\bibfnamefont {Jiangping}\ \bibnamefont {Hu}},\ }\bibfield  {title}
  {\enquote {\bibinfo {title} {Topological vortex phase transitions in
  iron-based superconductors},}\ }\href {\doibase
  https://doi.org/10.1016/j.scib.2019.07.011} {\bibfield  {journal} {\bibinfo
  {journal} {Science Bulletin}\ }\textbf {\bibinfo {volume} {64}},\ \bibinfo
  {pages} {1207 -- 1214} (\bibinfo {year} {2019})}\BibitemShut {NoStop}%
\bibitem [{\citenamefont {Yan}\ \emph {et~al.}(2020)\citenamefont {Yan},
  \citenamefont {Wu},\ and\ \citenamefont {Huang}}]{Yan2020vortex}%
  \BibitemOpen
  \bibfield  {author} {\bibinfo {author} {\bibfnamefont {Zhongbo}\ \bibnamefont
  {Yan}}, \bibinfo {author} {\bibfnamefont {Zhigang}\ \bibnamefont {Wu}}, \
  and\ \bibinfo {author} {\bibfnamefont {Wen}\ \bibnamefont {Huang}},\
  }\bibfield  {title} {\enquote {\bibinfo {title} {Vortex end majorana zero
  modes in superconducting dirac and weyl semimetals},}\ }\href {\doibase
  10.1103/PhysRevLett.124.257001} {\bibfield  {journal} {\bibinfo  {journal}
  {Phys. Rev. Lett.}\ }\textbf {\bibinfo {volume} {124}},\ \bibinfo {pages}
  {257001} (\bibinfo {year} {2020})}\BibitemShut {NoStop}%
\bibitem [{\citenamefont {Yan}\ \emph {et~al.}(2018)\citenamefont {Yan},
  \citenamefont {Song},\ and\ \citenamefont {Wang}}]{Yan2018hosc}%
  \BibitemOpen
  \bibfield  {author} {\bibinfo {author} {\bibfnamefont {Zhongbo}\ \bibnamefont
  {Yan}}, \bibinfo {author} {\bibfnamefont {Fei}\ \bibnamefont {Song}}, \ and\
  \bibinfo {author} {\bibfnamefont {Zhong}\ \bibnamefont {Wang}},\ }\bibfield
  {title} {\enquote {\bibinfo {title} {Majorana corner modes in a
  high-temperature platform},}\ }\href {\doibase
  10.1103/PhysRevLett.121.096803} {\bibfield  {journal} {\bibinfo  {journal}
  {Phys. Rev. Lett.}\ }\textbf {\bibinfo {volume} {121}},\ \bibinfo {pages}
  {096803} (\bibinfo {year} {2018})}\BibitemShut {NoStop}%
\bibitem [{\citenamefont {Wang}\ \emph
  {et~al.}(2018{\natexlab{a}})\citenamefont {Wang}, \citenamefont {Liu},
  \citenamefont {Lu},\ and\ \citenamefont {Zhang}}]{Wang2018hosc}%
  \BibitemOpen
  \bibfield  {author} {\bibinfo {author} {\bibfnamefont {Qiyue}\ \bibnamefont
  {Wang}}, \bibinfo {author} {\bibfnamefont {Cheng-Cheng}\ \bibnamefont {Liu}},
  \bibinfo {author} {\bibfnamefont {Yuan-Ming}\ \bibnamefont {Lu}}, \ and\
  \bibinfo {author} {\bibfnamefont {Fan}\ \bibnamefont {Zhang}},\ }\bibfield
  {title} {\enquote {\bibinfo {title} {High-temperature majorana corner
  states},}\ }\href {\doibase 10.1103/PhysRevLett.121.186801} {\bibfield
  {journal} {\bibinfo  {journal} {Phys. Rev. Lett.}\ }\textbf {\bibinfo
  {volume} {121}},\ \bibinfo {pages} {186801} (\bibinfo {year}
  {2018}{\natexlab{a}})}\BibitemShut {NoStop}%
\bibitem [{\citenamefont {Zhang}\ \emph
  {et~al.}(2019{\natexlab{a}})\citenamefont {Zhang}, \citenamefont {Cole},\
  and\ \citenamefont {Das~Sarma}}]{Zhang2019hinge}%
  \BibitemOpen
  \bibfield  {author} {\bibinfo {author} {\bibfnamefont {Rui-Xing}\
  \bibnamefont {Zhang}}, \bibinfo {author} {\bibfnamefont {William~S.}\
  \bibnamefont {Cole}}, \ and\ \bibinfo {author} {\bibfnamefont
  {S.}~\bibnamefont {Das~Sarma}},\ }\bibfield  {title} {\enquote {\bibinfo
  {title} {Helical hinge majorana modes in iron-based superconductors},}\
  }\href {\doibase 10.1103/PhysRevLett.122.187001} {\bibfield  {journal}
  {\bibinfo  {journal} {Phys. Rev. Lett.}\ }\textbf {\bibinfo {volume} {122}},\
  \bibinfo {pages} {187001} (\bibinfo {year} {2019}{\natexlab{a}})}\BibitemShut
  {NoStop}%
\bibitem [{\citenamefont {Gray}\ \emph {et~al.}(2019)\citenamefont {Gray},
  \citenamefont {Freudenstein}, \citenamefont {Zhao}, \citenamefont
  {O’Connor}, \citenamefont {Jenkins}, \citenamefont {Kumar}, \citenamefont
  {Hoek}, \citenamefont {Kopec}, \citenamefont {Huh}, \citenamefont
  {Taniguchi}, \citenamefont {Watanabe}, \citenamefont {Zhong}, \citenamefont
  {Kim}, \citenamefont {Gu},\ and\ \citenamefont {Burch}}]{Gray2019helical}%
  \BibitemOpen
  \bibfield  {author} {\bibinfo {author} {\bibfnamefont {Mason~J.}\
  \bibnamefont {Gray}}, \bibinfo {author} {\bibfnamefont {Josef}\ \bibnamefont
  {Freudenstein}}, \bibinfo {author} {\bibfnamefont {Shu Yang~F.}\ \bibnamefont
  {Zhao}}, \bibinfo {author} {\bibfnamefont {Ryan}\ \bibnamefont {O’Connor}},
  \bibinfo {author} {\bibfnamefont {Samuel}\ \bibnamefont {Jenkins}}, \bibinfo
  {author} {\bibfnamefont {Narendra}\ \bibnamefont {Kumar}}, \bibinfo {author}
  {\bibfnamefont {Marcel}\ \bibnamefont {Hoek}}, \bibinfo {author}
  {\bibfnamefont {Abigail}\ \bibnamefont {Kopec}}, \bibinfo {author}
  {\bibfnamefont {Soonsang}\ \bibnamefont {Huh}}, \bibinfo {author}
  {\bibfnamefont {Takashi}\ \bibnamefont {Taniguchi}}, \bibinfo {author}
  {\bibfnamefont {Kenji}\ \bibnamefont {Watanabe}}, \bibinfo {author}
  {\bibfnamefont {Ruidan}\ \bibnamefont {Zhong}}, \bibinfo {author}
  {\bibfnamefont {Changyoung}\ \bibnamefont {Kim}}, \bibinfo {author}
  {\bibfnamefont {G.~D.}\ \bibnamefont {Gu}}, \ and\ \bibinfo {author}
  {\bibfnamefont {K.~S.}\ \bibnamefont {Burch}},\ }\bibfield  {title} {\enquote
  {\bibinfo {title} {Evidence for helical hinge zero modes in an fe-based
  superconductor},}\ }\href {\doibase 10.1021/acs.nanolett.9b00844} {\bibfield
  {journal} {\bibinfo  {journal} {Nano Letters}\ }\textbf {\bibinfo {volume}
  {19}},\ \bibinfo {pages} {4890--4896} (\bibinfo {year} {2019})}\BibitemShut
  {NoStop}%
\bibitem [{\citenamefont {Hsu}\ \emph {et~al.}(2020)\citenamefont {Hsu},
  \citenamefont {Cole}, \citenamefont {Zhang},\ and\ \citenamefont
  {Sau}}]{Hsu2020hosc}%
  \BibitemOpen
  \bibfield  {author} {\bibinfo {author} {\bibfnamefont {Yi-Ting}\ \bibnamefont
  {Hsu}}, \bibinfo {author} {\bibfnamefont {William~S.}\ \bibnamefont {Cole}},
  \bibinfo {author} {\bibfnamefont {Rui-Xing}\ \bibnamefont {Zhang}}, \ and\
  \bibinfo {author} {\bibfnamefont {Jay~D.}\ \bibnamefont {Sau}},\ }\bibfield
  {title} {\enquote {\bibinfo {title} {Inversion-protected higher-order
  topological superconductivity in monolayer ${\mathrm{wte}}_{2}$},}\ }\href
  {\doibase 10.1103/PhysRevLett.125.097001} {\bibfield  {journal} {\bibinfo
  {journal} {Phys. Rev. Lett.}\ }\textbf {\bibinfo {volume} {125}},\ \bibinfo
  {pages} {097001} (\bibinfo {year} {2020})}\BibitemShut {NoStop}%
\bibitem [{\citenamefont {Vu}\ \emph {et~al.}(2020)\citenamefont {Vu},
  \citenamefont {Zhang},\ and\ \citenamefont
  {Sarma}}]{vu2020timereversalinvariant}%
  \BibitemOpen
  \bibfield  {author} {\bibinfo {author} {\bibfnamefont {DinhDuy}\ \bibnamefont
  {Vu}}, \bibinfo {author} {\bibfnamefont {Rui-Xing}\ \bibnamefont {Zhang}}, \
  and\ \bibinfo {author} {\bibfnamefont {Sankar~Das}\ \bibnamefont {Sarma}},\
  }\href@noop {} {\enquote {\bibinfo {title} {Time-reversal-invariant
  $c_2$-symmetric higher-order topological superconductors},}\ } (\bibinfo
  {year} {2020}),\ \Eprint {http://arxiv.org/abs/2005.03679} {arXiv:2005.03679
  [cond-mat.supr-con]} \BibitemShut {NoStop}%
\bibitem [{\citenamefont {Wu}\ \emph {et~al.}(2020{\natexlab{a}})\citenamefont
  {Wu}, \citenamefont {Guo}, \citenamefont {Zheng},\ and\ \citenamefont
  {Zou}}]{wu2020boundaryobstructeda}%
  \BibitemOpen
  \bibfield  {author} {\bibinfo {author} {\bibfnamefont {Yu-Biao}\ \bibnamefont
  {Wu}}, \bibinfo {author} {\bibfnamefont {Guang-Can}\ \bibnamefont {Guo}},
  \bibinfo {author} {\bibfnamefont {Zhen}\ \bibnamefont {Zheng}}, \ and\
  \bibinfo {author} {\bibfnamefont {Xu-Bo}\ \bibnamefont {Zou}},\ }\href@noop
  {} {\enquote {\bibinfo {title} {Boundary-obstructed topological superfluids
  in staggered spin-orbit coupled fermi gases},}\ } (\bibinfo {year}
  {2020}{\natexlab{a}}),\ \Eprint {http://arxiv.org/abs/2007.15886}
  {arXiv:2007.15886 [cond-mat.quant-gas]} \BibitemShut {NoStop}%
\bibitem [{\citenamefont {Yan}(2019{\natexlab{a}})}]{Yan2019hosca}%
  \BibitemOpen
  \bibfield  {author} {\bibinfo {author} {\bibfnamefont {Zhongbo}\ \bibnamefont
  {Yan}},\ }\bibfield  {title} {\enquote {\bibinfo {title} {Higher-order
  topological odd-parity superconductors},}\ }\href {\doibase
  10.1103/PhysRevLett.123.177001} {\bibfield  {journal} {\bibinfo  {journal}
  {Phys. Rev. Lett.}\ }\textbf {\bibinfo {volume} {123}},\ \bibinfo {pages}
  {177001} (\bibinfo {year} {2019}{\natexlab{a}})}\BibitemShut {NoStop}%
\bibitem [{\citenamefont {Ahn}\ and\ \citenamefont {Yang}(2020)}]{Ahn2020hosc}%
  \BibitemOpen
  \bibfield  {author} {\bibinfo {author} {\bibfnamefont {Junyeong}\
  \bibnamefont {Ahn}}\ and\ \bibinfo {author} {\bibfnamefont {Bohm-Jung}\
  \bibnamefont {Yang}},\ }\bibfield  {title} {\enquote {\bibinfo {title}
  {Higher-order topological superconductivity of spin-polarized fermions},}\
  }\href {\doibase 10.1103/PhysRevResearch.2.012060} {\bibfield  {journal}
  {\bibinfo  {journal} {Phys. Rev. Research}\ }\textbf {\bibinfo {volume}
  {2}},\ \bibinfo {pages} {012060} (\bibinfo {year} {2020})}\BibitemShut
  {NoStop}%
\bibitem [{\citenamefont {Langbehn}\ \emph {et~al.}(2017)\citenamefont
  {Langbehn}, \citenamefont {Peng}, \citenamefont {Trifunovic}, \citenamefont
  {von Oppen},\ and\ \citenamefont {Brouwer}}]{Langbehn2017}%
  \BibitemOpen
  \bibfield  {author} {\bibinfo {author} {\bibfnamefont {Josias}\ \bibnamefont
  {Langbehn}}, \bibinfo {author} {\bibfnamefont {Yang}\ \bibnamefont {Peng}},
  \bibinfo {author} {\bibfnamefont {Luka}\ \bibnamefont {Trifunovic}}, \bibinfo
  {author} {\bibfnamefont {Felix}\ \bibnamefont {von Oppen}}, \ and\ \bibinfo
  {author} {\bibfnamefont {Piet~W.}\ \bibnamefont {Brouwer}},\ }\bibfield
  {title} {\enquote {\bibinfo {title} {Reflection-symmetric second-order
  topological insulators and superconductors},}\ }\href {\doibase
  10.1103/PhysRevLett.119.246401} {\bibfield  {journal} {\bibinfo  {journal}
  {Phys. Rev. Lett.}\ }\textbf {\bibinfo {volume} {119}},\ \bibinfo {pages}
  {246401} (\bibinfo {year} {2017})}\BibitemShut {NoStop}%
\bibitem [{\citenamefont {Shapourian}\ \emph {et~al.}(2018)\citenamefont
  {Shapourian}, \citenamefont {Wang},\ and\ \citenamefont
  {Ryu}}]{Shapourian2018SOTSC}%
  \BibitemOpen
  \bibfield  {author} {\bibinfo {author} {\bibfnamefont {Hassan}\ \bibnamefont
  {Shapourian}}, \bibinfo {author} {\bibfnamefont {Yuxuan}\ \bibnamefont
  {Wang}}, \ and\ \bibinfo {author} {\bibfnamefont {Shinsei}\ \bibnamefont
  {Ryu}},\ }\bibfield  {title} {\enquote {\bibinfo {title} {Topological
  crystalline superconductivity and second-order topological superconductivity
  in nodal-loop materials},}\ }\href {\doibase 10.1103/PhysRevB.97.094508}
  {\bibfield  {journal} {\bibinfo  {journal} {Phys. Rev. B}\ }\textbf {\bibinfo
  {volume} {97}},\ \bibinfo {pages} {094508} (\bibinfo {year}
  {2018})}\BibitemShut {NoStop}%
\bibitem [{\citenamefont {Khalaf}(2018)}]{Khalaf2018}%
  \BibitemOpen
  \bibfield  {author} {\bibinfo {author} {\bibfnamefont {Eslam}\ \bibnamefont
  {Khalaf}},\ }\bibfield  {title} {\enquote {\bibinfo {title} {Higher-order
  topological insulators and superconductors protected by inversion
  symmetry},}\ }\href {\doibase 10.1103/PhysRevB.97.205136} {\bibfield
  {journal} {\bibinfo  {journal} {Phys. Rev. B}\ }\textbf {\bibinfo {volume}
  {97}},\ \bibinfo {pages} {205136} (\bibinfo {year} {2018})}\BibitemShut
  {NoStop}%
\bibitem [{\citenamefont {Geier}\ \emph {et~al.}(2018)\citenamefont {Geier},
  \citenamefont {Trifunovic}, \citenamefont {Hoskam},\ and\ \citenamefont
  {Brouwer}}]{Geier2018}%
  \BibitemOpen
  \bibfield  {author} {\bibinfo {author} {\bibfnamefont {Max}\ \bibnamefont
  {Geier}}, \bibinfo {author} {\bibfnamefont {Luka}\ \bibnamefont
  {Trifunovic}}, \bibinfo {author} {\bibfnamefont {Max}\ \bibnamefont
  {Hoskam}}, \ and\ \bibinfo {author} {\bibfnamefont {Piet~W.}\ \bibnamefont
  {Brouwer}},\ }\bibfield  {title} {\enquote {\bibinfo {title} {Second-order
  topological insulators and superconductors with an order-two crystalline
  symmetry},}\ }\href {\doibase 10.1103/PhysRevB.97.205135} {\bibfield
  {journal} {\bibinfo  {journal} {Phys. Rev. B}\ }\textbf {\bibinfo {volume}
  {97}},\ \bibinfo {pages} {205135} (\bibinfo {year} {2018})}\BibitemShut
  {NoStop}%
\bibitem [{\citenamefont {Zhu}(2018)}]{Zhu2018hosc}%
  \BibitemOpen
  \bibfield  {author} {\bibinfo {author} {\bibfnamefont {Xiaoyu}\ \bibnamefont
  {Zhu}},\ }\bibfield  {title} {\enquote {\bibinfo {title} {Tunable majorana
  corner states in a two-dimensional second-order topological superconductor
  induced by magnetic fields},}\ }\href {\doibase 10.1103/PhysRevB.97.205134}
  {\bibfield  {journal} {\bibinfo  {journal} {Phys. Rev. B}\ }\textbf {\bibinfo
  {volume} {97}},\ \bibinfo {pages} {205134} (\bibinfo {year}
  {2018})}\BibitemShut {NoStop}%
\bibitem [{\citenamefont {Wang}\ \emph
  {et~al.}(2018{\natexlab{b}})\citenamefont {Wang}, \citenamefont {Lin},\ and\
  \citenamefont {Hughes}}]{Wangyuxuan2018hosc}%
  \BibitemOpen
  \bibfield  {author} {\bibinfo {author} {\bibfnamefont {Yuxuan}\ \bibnamefont
  {Wang}}, \bibinfo {author} {\bibfnamefont {Mao}\ \bibnamefont {Lin}}, \ and\
  \bibinfo {author} {\bibfnamefont {Taylor~L.}\ \bibnamefont {Hughes}},\
  }\bibfield  {title} {\enquote {\bibinfo {title} {Weak-pairing higher order
  topological superconductors},}\ }\href {\doibase 10.1103/PhysRevB.98.165144}
  {\bibfield  {journal} {\bibinfo  {journal} {Phys. Rev. B}\ }\textbf {\bibinfo
  {volume} {98}},\ \bibinfo {pages} {165144} (\bibinfo {year}
  {2018}{\natexlab{b}})}\BibitemShut {NoStop}%
\bibitem [{\citenamefont {Hsu}\ \emph {et~al.}(2018)\citenamefont {Hsu},
  \citenamefont {Stano}, \citenamefont {Klinovaja},\ and\ \citenamefont
  {Loss}}]{Hsu2018hosc}%
  \BibitemOpen
  \bibfield  {author} {\bibinfo {author} {\bibfnamefont {Chen-Hsuan}\
  \bibnamefont {Hsu}}, \bibinfo {author} {\bibfnamefont {Peter}\ \bibnamefont
  {Stano}}, \bibinfo {author} {\bibfnamefont {Jelena}\ \bibnamefont
  {Klinovaja}}, \ and\ \bibinfo {author} {\bibfnamefont {Daniel}\ \bibnamefont
  {Loss}},\ }\bibfield  {title} {\enquote {\bibinfo {title} {Majorana kramers
  pairs in higher-order topological insulators},}\ }\href {\doibase
  10.1103/PhysRevLett.121.196801} {\bibfield  {journal} {\bibinfo  {journal}
  {Phys. Rev. Lett.}\ }\textbf {\bibinfo {volume} {121}},\ \bibinfo {pages}
  {196801} (\bibinfo {year} {2018})}\BibitemShut {NoStop}%
\bibitem [{\citenamefont {Liu}\ \emph {et~al.}(2018{\natexlab{a}})\citenamefont
  {Liu}, \citenamefont {He},\ and\ \citenamefont {Nori}}]{Liu2018hosc}%
  \BibitemOpen
  \bibfield  {author} {\bibinfo {author} {\bibfnamefont {Tao}\ \bibnamefont
  {Liu}}, \bibinfo {author} {\bibfnamefont {James~Jun}\ \bibnamefont {He}}, \
  and\ \bibinfo {author} {\bibfnamefont {Franco}\ \bibnamefont {Nori}},\
  }\bibfield  {title} {\enquote {\bibinfo {title} {Majorana corner states in a
  two-dimensional magnetic topological insulator on a high-temperature
  superconductor},}\ }\href {\doibase 10.1103/PhysRevB.98.245413} {\bibfield
  {journal} {\bibinfo  {journal} {Phys. Rev. B}\ }\textbf {\bibinfo {volume}
  {98}},\ \bibinfo {pages} {245413} (\bibinfo {year}
  {2018}{\natexlab{a}})}\BibitemShut {NoStop}%
\bibitem [{\citenamefont {Wu}\ \emph {et~al.}(2019)\citenamefont {Wu},
  \citenamefont {Yan},\ and\ \citenamefont {Huang}}]{Wuzhigang2019hosc}%
  \BibitemOpen
  \bibfield  {author} {\bibinfo {author} {\bibfnamefont {Zhigang}\ \bibnamefont
  {Wu}}, \bibinfo {author} {\bibfnamefont {Zhongbo}\ \bibnamefont {Yan}}, \
  and\ \bibinfo {author} {\bibfnamefont {Wen}\ \bibnamefont {Huang}},\
  }\bibfield  {title} {\enquote {\bibinfo {title} {Higher-order topological
  superconductivity: Possible realization in fermi gases and
  ${\mathrm{sr}}_{2}{\mathrm{ruo}}_{4}$},}\ }\href {\doibase
  10.1103/PhysRevB.99.020508} {\bibfield  {journal} {\bibinfo  {journal} {Phys.
  Rev. B}\ }\textbf {\bibinfo {volume} {99}},\ \bibinfo {pages} {020508}
  (\bibinfo {year} {2019})}\BibitemShut {NoStop}%
\bibitem [{\citenamefont {Volpez}\ \emph {et~al.}(2019)\citenamefont {Volpez},
  \citenamefont {Loss},\ and\ \citenamefont {Klinovaja}}]{Volpez2019SOTSC}%
  \BibitemOpen
  \bibfield  {author} {\bibinfo {author} {\bibfnamefont {Yanick}\ \bibnamefont
  {Volpez}}, \bibinfo {author} {\bibfnamefont {Daniel}\ \bibnamefont {Loss}}, \
  and\ \bibinfo {author} {\bibfnamefont {Jelena}\ \bibnamefont {Klinovaja}},\
  }\bibfield  {title} {\enquote {\bibinfo {title} {Second-order topological
  superconductivity in $\ensuremath{\pi}$-junction rashba layers},}\ }\href
  {\doibase 10.1103/PhysRevLett.122.126402} {\bibfield  {journal} {\bibinfo
  {journal} {Phys. Rev. Lett.}\ }\textbf {\bibinfo {volume} {122}},\ \bibinfo
  {pages} {126402} (\bibinfo {year} {2019})}\BibitemShut {NoStop}%
\bibitem [{\citenamefont {Zhang}\ \emph
  {et~al.}(2019{\natexlab{b}})\citenamefont {Zhang}, \citenamefont {Cole},
  \citenamefont {Wu},\ and\ \citenamefont {Das~Sarma}}]{Zhang2019hoscb}%
  \BibitemOpen
  \bibfield  {author} {\bibinfo {author} {\bibfnamefont {Rui-Xing}\
  \bibnamefont {Zhang}}, \bibinfo {author} {\bibfnamefont {William~S.}\
  \bibnamefont {Cole}}, \bibinfo {author} {\bibfnamefont {Xianxin}\
  \bibnamefont {Wu}}, \ and\ \bibinfo {author} {\bibfnamefont {S.}~\bibnamefont
  {Das~Sarma}},\ }\bibfield  {title} {\enquote {\bibinfo {title} {Higher-order
  topology and nodal topological superconductivity in fe(se,te)
  heterostructures},}\ }\href {\doibase 10.1103/PhysRevLett.123.167001}
  {\bibfield  {journal} {\bibinfo  {journal} {Phys. Rev. Lett.}\ }\textbf
  {\bibinfo {volume} {123}},\ \bibinfo {pages} {167001} (\bibinfo {year}
  {2019}{\natexlab{b}})}\BibitemShut {NoStop}%
\bibitem [{\citenamefont {{Wu}}\ \emph {et~al.}(2019)\citenamefont {{Wu}},
  \citenamefont {{Liu}}, \citenamefont {{Thomale}},\ and\ \citenamefont
  {{Liu}}}]{Wu2019hoscb}%
  \BibitemOpen
  \bibfield  {author} {\bibinfo {author} {\bibfnamefont {Xianxin}\ \bibnamefont
  {{Wu}}}, \bibinfo {author} {\bibfnamefont {Xin}\ \bibnamefont {{Liu}}},
  \bibinfo {author} {\bibfnamefont {Ronny}\ \bibnamefont {{Thomale}}}, \ and\
  \bibinfo {author} {\bibfnamefont {Chao-Xing}\ \bibnamefont {{Liu}}},\
  }\bibfield  {title} {\enquote {\bibinfo {title} {{High-$T_c$ Superconductor
  Fe(Se,Te) Monolayer: an Intrinsic, Scalable and Electrically-tunable Majorana
  Platform}},}\ }\href@noop {} {\bibfield  {journal} {\bibinfo  {journal}
  {arXiv e-prints}\ ,\ \bibinfo {eid} {arXiv:1905.10648}} (\bibinfo {year}
  {2019})},\ \Eprint {http://arxiv.org/abs/1905.10648} {arXiv:1905.10648
  [cond-mat.supr-con]} \BibitemShut {NoStop}%
\bibitem [{\citenamefont {Zeng}\ \emph {et~al.}(2019)\citenamefont {Zeng},
  \citenamefont {Stanescu}, \citenamefont {Zhang}, \citenamefont {Scarola},\
  and\ \citenamefont {Tewari}}]{Zeng2019mcm}%
  \BibitemOpen
  \bibfield  {author} {\bibinfo {author} {\bibfnamefont {Chuanchang}\
  \bibnamefont {Zeng}}, \bibinfo {author} {\bibfnamefont {T.~D.}\ \bibnamefont
  {Stanescu}}, \bibinfo {author} {\bibfnamefont {Chuanwei}\ \bibnamefont
  {Zhang}}, \bibinfo {author} {\bibfnamefont {V.~W.}\ \bibnamefont {Scarola}},
  \ and\ \bibinfo {author} {\bibfnamefont {Sumanta}\ \bibnamefont {Tewari}},\
  }\bibfield  {title} {\enquote {\bibinfo {title} {Majorana corner modes with
  solitons in an attractive hubbard-hofstadter model of cold atom optical
  lattices},}\ }\href {\doibase 10.1103/PhysRevLett.123.060402} {\bibfield
  {journal} {\bibinfo  {journal} {Phys. Rev. Lett.}\ }\textbf {\bibinfo
  {volume} {123}},\ \bibinfo {pages} {060402} (\bibinfo {year}
  {2019})}\BibitemShut {NoStop}%
\bibitem [{\citenamefont {Bultinck}\ \emph {et~al.}(2019)\citenamefont
  {Bultinck}, \citenamefont {Bernevig},\ and\ \citenamefont
  {Zaletel}}]{Bultinck2019}%
  \BibitemOpen
  \bibfield  {author} {\bibinfo {author} {\bibfnamefont {Nick}\ \bibnamefont
  {Bultinck}}, \bibinfo {author} {\bibfnamefont {B.~Andrei}\ \bibnamefont
  {Bernevig}}, \ and\ \bibinfo {author} {\bibfnamefont {Michael~P.}\
  \bibnamefont {Zaletel}},\ }\bibfield  {title} {\enquote {\bibinfo {title}
  {Three-dimensional superconductors with hybrid higher-order topology},}\
  }\href {\doibase 10.1103/PhysRevB.99.125149} {\bibfield  {journal} {\bibinfo
  {journal} {Phys. Rev. B}\ }\textbf {\bibinfo {volume} {99}},\ \bibinfo
  {pages} {125149} (\bibinfo {year} {2019})}\BibitemShut {NoStop}%
\bibitem [{\citenamefont {Ghorashi}\ \emph {et~al.}(2019)\citenamefont
  {Ghorashi}, \citenamefont {Hu}, \citenamefont {Hughes},\ and\ \citenamefont
  {Rossi}}]{Ghorashi2019}%
  \BibitemOpen
  \bibfield  {author} {\bibinfo {author} {\bibfnamefont {Sayed Ali~Akbar}\
  \bibnamefont {Ghorashi}}, \bibinfo {author} {\bibfnamefont {Xiang}\
  \bibnamefont {Hu}}, \bibinfo {author} {\bibfnamefont {Taylor~L.}\
  \bibnamefont {Hughes}}, \ and\ \bibinfo {author} {\bibfnamefont {Enrico}\
  \bibnamefont {Rossi}},\ }\bibfield  {title} {\enquote {\bibinfo {title}
  {Second-order dirac superconductors and magnetic field induced majorana hinge
  modes},}\ }\href {\doibase 10.1103/PhysRevB.100.020509} {\bibfield  {journal}
  {\bibinfo  {journal} {Phys. Rev. B}\ }\textbf {\bibinfo {volume} {100}},\
  \bibinfo {pages} {020509} (\bibinfo {year} {2019})}\BibitemShut {NoStop}%
\bibitem [{\citenamefont {Peng}\ and\ \citenamefont
  {Xu}(2019)}]{Peng2019hinge}%
  \BibitemOpen
  \bibfield  {author} {\bibinfo {author} {\bibfnamefont {Yang}\ \bibnamefont
  {Peng}}\ and\ \bibinfo {author} {\bibfnamefont {Yong}\ \bibnamefont {Xu}},\
  }\bibfield  {title} {\enquote {\bibinfo {title} {Proximity-induced majorana
  hinge modes in antiferromagnetic topological insulators},}\ }\href {\doibase
  10.1103/PhysRevB.99.195431} {\bibfield  {journal} {\bibinfo  {journal} {Phys.
  Rev. B}\ }\textbf {\bibinfo {volume} {99}},\ \bibinfo {pages} {195431}
  (\bibinfo {year} {2019})}\BibitemShut {NoStop}%
\bibitem [{\citenamefont {Zhu}(2019)}]{Zhu2019mixed}%
  \BibitemOpen
  \bibfield  {author} {\bibinfo {author} {\bibfnamefont {Xiaoyu}\ \bibnamefont
  {Zhu}},\ }\bibfield  {title} {\enquote {\bibinfo {title} {Second-order
  topological superconductors with mixed pairing},}\ }\href {\doibase
  10.1103/PhysRevLett.122.236401} {\bibfield  {journal} {\bibinfo  {journal}
  {Phys. Rev. Lett.}\ }\textbf {\bibinfo {volume} {122}},\ \bibinfo {pages}
  {236401} (\bibinfo {year} {2019})}\BibitemShut {NoStop}%
\bibitem [{\citenamefont {Laubscher}\ \emph {et~al.}(2019)\citenamefont
  {Laubscher}, \citenamefont {Loss},\ and\ \citenamefont
  {Klinovaja}}]{Laubscher2019hosc}%
  \BibitemOpen
  \bibfield  {author} {\bibinfo {author} {\bibfnamefont {Katharina}\
  \bibnamefont {Laubscher}}, \bibinfo {author} {\bibfnamefont {Daniel}\
  \bibnamefont {Loss}}, \ and\ \bibinfo {author} {\bibfnamefont {Jelena}\
  \bibnamefont {Klinovaja}},\ }\bibfield  {title} {\enquote {\bibinfo {title}
  {Fractional topological superconductivity and parafermion corner states},}\
  }\href {\doibase 10.1103/PhysRevResearch.1.032017} {\bibfield  {journal}
  {\bibinfo  {journal} {Phys. Rev. Research}\ }\textbf {\bibinfo {volume}
  {1}},\ \bibinfo {pages} {032017} (\bibinfo {year} {2019})}\BibitemShut
  {NoStop}%
\bibitem [{\citenamefont {Pan}\ \emph {et~al.}(2019)\citenamefont {Pan},
  \citenamefont {Yang}, \citenamefont {Chen}, \citenamefont {Xu}, \citenamefont
  {Liu},\ and\ \citenamefont {Liu}}]{Pan2019SOTSC}%
  \BibitemOpen
  \bibfield  {author} {\bibinfo {author} {\bibfnamefont {Xiao-Hong}\
  \bibnamefont {Pan}}, \bibinfo {author} {\bibfnamefont {Kai-Jie}\ \bibnamefont
  {Yang}}, \bibinfo {author} {\bibfnamefont {Li}~\bibnamefont {Chen}}, \bibinfo
  {author} {\bibfnamefont {Gang}\ \bibnamefont {Xu}}, \bibinfo {author}
  {\bibfnamefont {Chao-Xing}\ \bibnamefont {Liu}}, \ and\ \bibinfo {author}
  {\bibfnamefont {Xin}\ \bibnamefont {Liu}},\ }\bibfield  {title} {\enquote
  {\bibinfo {title} {Lattice-symmetry-assisted second-order topological
  superconductors and majorana patterns},}\ }\href {\doibase
  10.1103/PhysRevLett.123.156801} {\bibfield  {journal} {\bibinfo  {journal}
  {Phys. Rev. Lett.}\ }\textbf {\bibinfo {volume} {123}},\ \bibinfo {pages}
  {156801} (\bibinfo {year} {2019})}\BibitemShut {NoStop}%
\bibitem [{\citenamefont {Yan}(2019{\natexlab{b}})}]{Yan2019hoscb}%
  \BibitemOpen
  \bibfield  {author} {\bibinfo {author} {\bibfnamefont {Zhongbo}\ \bibnamefont
  {Yan}},\ }\bibfield  {title} {\enquote {\bibinfo {title} {Majorana corner and
  hinge modes in second-order topological insulator/superconductor
  heterostructures},}\ }\href {\doibase 10.1103/PhysRevB.100.205406} {\bibfield
   {journal} {\bibinfo  {journal} {Phys. Rev. B}\ }\textbf {\bibinfo {volume}
  {100}},\ \bibinfo {pages} {205406} (\bibinfo {year}
  {2019}{\natexlab{b}})}\BibitemShut {NoStop}%
\bibitem [{\citenamefont {Franca}\ \emph {et~al.}(2019)\citenamefont {Franca},
  \citenamefont {Efremov},\ and\ \citenamefont {Fulga}}]{Franca2019SOTSC}%
  \BibitemOpen
  \bibfield  {author} {\bibinfo {author} {\bibfnamefont {S.}~\bibnamefont
  {Franca}}, \bibinfo {author} {\bibfnamefont {D.~V.}\ \bibnamefont {Efremov}},
  \ and\ \bibinfo {author} {\bibfnamefont {I.~C.}\ \bibnamefont {Fulga}},\
  }\bibfield  {title} {\enquote {\bibinfo {title} {Phase-tunable second-order
  topological superconductor},}\ }\href {\doibase 10.1103/PhysRevB.100.075415}
  {\bibfield  {journal} {\bibinfo  {journal} {Phys. Rev. B}\ }\textbf {\bibinfo
  {volume} {100}},\ \bibinfo {pages} {075415} (\bibinfo {year}
  {2019})}\BibitemShut {NoStop}%
\bibitem [{\citenamefont {Kheirkhah}\ \emph
  {et~al.}(2020{\natexlab{a}})\citenamefont {Kheirkhah}, \citenamefont {Nagai},
  \citenamefont {Chen},\ and\ \citenamefont {Marsiglio}}]{Majid2020hosca}%
  \BibitemOpen
  \bibfield  {author} {\bibinfo {author} {\bibfnamefont {Majid}\ \bibnamefont
  {Kheirkhah}}, \bibinfo {author} {\bibfnamefont {Yuki}\ \bibnamefont {Nagai}},
  \bibinfo {author} {\bibfnamefont {Chun}\ \bibnamefont {Chen}}, \ and\
  \bibinfo {author} {\bibfnamefont {Frank}\ \bibnamefont {Marsiglio}},\
  }\bibfield  {title} {\enquote {\bibinfo {title} {Majorana corner flat bands
  in two-dimensional second-order topological superconductors},}\ }\href
  {\doibase 10.1103/PhysRevB.101.104502} {\bibfield  {journal} {\bibinfo
  {journal} {Phys. Rev. B}\ }\textbf {\bibinfo {volume} {101}},\ \bibinfo
  {pages} {104502} (\bibinfo {year} {2020}{\natexlab{a}})}\BibitemShut
  {NoStop}%
\bibitem [{\citenamefont {Zhang}\ and\ \citenamefont
  {Trauzettel}(2020)}]{Zhang2020SOTSC}%
  \BibitemOpen
  \bibfield  {author} {\bibinfo {author} {\bibfnamefont {Song-Bo}\ \bibnamefont
  {Zhang}}\ and\ \bibinfo {author} {\bibfnamefont {Bj\"orn}\ \bibnamefont
  {Trauzettel}},\ }\bibfield  {title} {\enquote {\bibinfo {title} {Detection of
  second-order topological superconductors by josephson junctions},}\ }\href
  {\doibase 10.1103/PhysRevResearch.2.012018} {\bibfield  {journal} {\bibinfo
  {journal} {Phys. Rev. Research}\ }\textbf {\bibinfo {volume} {2}},\ \bibinfo
  {pages} {012018} (\bibinfo {year} {2020})}\BibitemShut {NoStop}%
\bibitem [{\citenamefont {Roy}(2020)}]{Bitan2020hosc}%
  \BibitemOpen
  \bibfield  {author} {\bibinfo {author} {\bibfnamefont {Bitan}\ \bibnamefont
  {Roy}},\ }\bibfield  {title} {\enquote {\bibinfo {title} {Higher-order
  topological superconductors in $\mathcal{P}$-, $\mathcal{T}$-odd quadrupolar
  dirac materials},}\ }\href {\doibase 10.1103/PhysRevB.101.220506} {\bibfield
  {journal} {\bibinfo  {journal} {Phys. Rev. B}\ }\textbf {\bibinfo {volume}
  {101}},\ \bibinfo {pages} {220506} (\bibinfo {year} {2020})}\BibitemShut
  {NoStop}%
\bibitem [{\citenamefont {Wu}\ \emph {et~al.}(2020{\natexlab{b}})\citenamefont
  {Wu}, \citenamefont {Hou}, \citenamefont {Li}, \citenamefont {Luo},
  \citenamefont {Shi},\ and\ \citenamefont {Zhang}}]{Wu2020SOTSC}%
  \BibitemOpen
  \bibfield  {author} {\bibinfo {author} {\bibfnamefont {Ya-Jie}\ \bibnamefont
  {Wu}}, \bibinfo {author} {\bibfnamefont {Junpeng}\ \bibnamefont {Hou}},
  \bibinfo {author} {\bibfnamefont {Yun-Mei}\ \bibnamefont {Li}}, \bibinfo
  {author} {\bibfnamefont {Xi-Wang}\ \bibnamefont {Luo}}, \bibinfo {author}
  {\bibfnamefont {Xiaoyan}\ \bibnamefont {Shi}}, \ and\ \bibinfo {author}
  {\bibfnamefont {Chuanwei}\ \bibnamefont {Zhang}},\ }\bibfield  {title}
  {\enquote {\bibinfo {title} {In-plane zeeman-field-induced majorana corner
  and hinge modes in an $s$-wave superconductor heterostructure},}\ }\href
  {\doibase 10.1103/PhysRevLett.124.227001} {\bibfield  {journal} {\bibinfo
  {journal} {Phys. Rev. Lett.}\ }\textbf {\bibinfo {volume} {124}},\ \bibinfo
  {pages} {227001} (\bibinfo {year} {2020}{\natexlab{b}})}\BibitemShut
  {NoStop}%
\bibitem [{\citenamefont {Kheirkhah}\ \emph
  {et~al.}(2020{\natexlab{b}})\citenamefont {Kheirkhah}, \citenamefont {Yan},
  \citenamefont {Nagai},\ and\ \citenamefont {Marsiglio}}]{Majid2020hoscb}%
  \BibitemOpen
  \bibfield  {author} {\bibinfo {author} {\bibfnamefont {Majid}\ \bibnamefont
  {Kheirkhah}}, \bibinfo {author} {\bibfnamefont {Zhongbo}\ \bibnamefont
  {Yan}}, \bibinfo {author} {\bibfnamefont {Yuki}\ \bibnamefont {Nagai}}, \
  and\ \bibinfo {author} {\bibfnamefont {Frank}\ \bibnamefont {Marsiglio}},\
  }\bibfield  {title} {\enquote {\bibinfo {title} {First- and second-order
  topological superconductivity and temperature-driven topological phase
  transitions in the extended hubbard model with spin-orbit coupling},}\ }\href
  {\doibase 10.1103/PhysRevLett.125.017001} {\bibfield  {journal} {\bibinfo
  {journal} {Phys. Rev. Lett.}\ }\textbf {\bibinfo {volume} {125}},\ \bibinfo
  {pages} {017001} (\bibinfo {year} {2020}{\natexlab{b}})}\BibitemShut
  {NoStop}%
\bibitem [{\citenamefont {Wu}\ \emph {et~al.}(2020{\natexlab{c}})\citenamefont
  {Wu}, \citenamefont {Benalcazar}, \citenamefont {Li}, \citenamefont
  {Thomale}, \citenamefont {Liu},\ and\ \citenamefont
  {Hu}}]{wu2020boundaryobstructedb}%
  \BibitemOpen
  \bibfield  {author} {\bibinfo {author} {\bibfnamefont {Xianxin}\ \bibnamefont
  {Wu}}, \bibinfo {author} {\bibfnamefont {Wladimir~A.}\ \bibnamefont
  {Benalcazar}}, \bibinfo {author} {\bibfnamefont {Yinxiang}\ \bibnamefont
  {Li}}, \bibinfo {author} {\bibfnamefont {Ronny}\ \bibnamefont {Thomale}},
  \bibinfo {author} {\bibfnamefont {Chao-Xing}\ \bibnamefont {Liu}}, \ and\
  \bibinfo {author} {\bibfnamefont {Jiangping}\ \bibnamefont {Hu}},\
  }\href@noop {} {\enquote {\bibinfo {title} {Boundary-obstructed topological
  high-t$_c$ superconductivity in iron pnictides},}\ } (\bibinfo {year}
  {2020}{\natexlab{c}}),\ \Eprint {http://arxiv.org/abs/2003.12204}
  {arXiv:2003.12204 [cond-mat.supr-con]} \BibitemShut {NoStop}%
\bibitem [{\citenamefont {Tiwari}\ \emph {et~al.}(2020)\citenamefont {Tiwari},
  \citenamefont {Jahin},\ and\ \citenamefont {Wang}}]{tiwari2020chiral}%
  \BibitemOpen
  \bibfield  {author} {\bibinfo {author} {\bibfnamefont {Apoorv}\ \bibnamefont
  {Tiwari}}, \bibinfo {author} {\bibfnamefont {Ammar}\ \bibnamefont {Jahin}}, \
  and\ \bibinfo {author} {\bibfnamefont {Yuxuan}\ \bibnamefont {Wang}},\
  }\href@noop {} {\enquote {\bibinfo {title} {Chiral dirac superconductors:
  Second-order and boundary-obstructed topology},}\ } (\bibinfo {year}
  {2020}),\ \Eprint {http://arxiv.org/abs/2005.12291} {arXiv:2005.12291
  [cond-mat.mes-hall]} \BibitemShut {NoStop}%
\bibitem [{\citenamefont {Ono}\ \emph {et~al.}(2020{\natexlab{a}})\citenamefont
  {Ono}, \citenamefont {Po},\ and\ \citenamefont {Watanabe}}]{Ono2020SI}%
  \BibitemOpen
  \bibfield  {author} {\bibinfo {author} {\bibfnamefont {Seishiro}\
  \bibnamefont {Ono}}, \bibinfo {author} {\bibfnamefont {Hoi~Chun}\
  \bibnamefont {Po}}, \ and\ \bibinfo {author} {\bibfnamefont {Haruki}\
  \bibnamefont {Watanabe}},\ }\bibfield  {title} {\enquote {\bibinfo {title}
  {Refined symmetry indicators for topological superconductors in all space
  groups},}\ }\href {\doibase 10.1126/sciadv.aaz8367} {\bibfield  {journal}
  {\bibinfo  {journal} {Science Advances}\ }\textbf {\bibinfo {volume} {6}}
  (\bibinfo {year} {2020}{\natexlab{a}}),\ 10.1126/sciadv.aaz8367},\ \Eprint
  {http://arxiv.org/abs/https://advances.sciencemag.org/content/6/18/eaaz8367.full.pdf}
  {https://advances.sciencemag.org/content/6/18/eaaz8367.full.pdf} \BibitemShut
  {NoStop}%
\bibitem [{\citenamefont {Ono}\ \emph {et~al.}(2020{\natexlab{b}})\citenamefont
  {Ono}, \citenamefont {Po},\ and\ \citenamefont {Shiozaki}}]{ono2020enriched}%
  \BibitemOpen
  \bibfield  {author} {\bibinfo {author} {\bibfnamefont {Seishiro}\
  \bibnamefont {Ono}}, \bibinfo {author} {\bibfnamefont {Hoi~Chun}\
  \bibnamefont {Po}}, \ and\ \bibinfo {author} {\bibfnamefont {Ken}\
  \bibnamefont {Shiozaki}},\ }\href@noop {} {\enquote {\bibinfo {title}
  {$z_2$-enriched symmetry indicators for topological superconductors in the
  1651 magnetic space groups},}\ } (\bibinfo {year} {2020}{\natexlab{b}}),\
  \Eprint {http://arxiv.org/abs/2008.05499} {arXiv:2008.05499
  [cond-mat.supr-con]} \BibitemShut {NoStop}%
\bibitem [{\citenamefont {Wang}\ \emph {et~al.}(2015)\citenamefont {Wang},
  \citenamefont {Zhang}, \citenamefont {Xu}, \citenamefont {Zeng},
  \citenamefont {Miao}, \citenamefont {Xu}, \citenamefont {Qian}, \citenamefont
  {Weng}, \citenamefont {Richard}, \citenamefont {Fedorov}, \citenamefont
  {Ding}, \citenamefont {Dai},\ and\ \citenamefont {Fang}}]{Wang2015iFeSC}%
  \BibitemOpen
  \bibfield  {author} {\bibinfo {author} {\bibfnamefont {Zhijun}\ \bibnamefont
  {Wang}}, \bibinfo {author} {\bibfnamefont {P.}~\bibnamefont {Zhang}},
  \bibinfo {author} {\bibfnamefont {Gang}\ \bibnamefont {Xu}}, \bibinfo
  {author} {\bibfnamefont {L.~K.}\ \bibnamefont {Zeng}}, \bibinfo {author}
  {\bibfnamefont {H.}~\bibnamefont {Miao}}, \bibinfo {author} {\bibfnamefont
  {Xiaoyan}\ \bibnamefont {Xu}}, \bibinfo {author} {\bibfnamefont
  {T.}~\bibnamefont {Qian}}, \bibinfo {author} {\bibfnamefont {Hongming}\
  \bibnamefont {Weng}}, \bibinfo {author} {\bibfnamefont {P.}~\bibnamefont
  {Richard}}, \bibinfo {author} {\bibfnamefont {A.~V.}\ \bibnamefont
  {Fedorov}}, \bibinfo {author} {\bibfnamefont {H.}~\bibnamefont {Ding}},
  \bibinfo {author} {\bibfnamefont {Xi}~\bibnamefont {Dai}}, \ and\ \bibinfo
  {author} {\bibfnamefont {Zhong}\ \bibnamefont {Fang}},\ }\bibfield  {title}
  {\enquote {\bibinfo {title} {Topological nature of the
  ${\mathrm{fese}}_{0.5}{\mathrm{te}}_{0.5}$ superconductor},}\ }\href
  {\doibase 10.1103/PhysRevB.92.115119} {\bibfield  {journal} {\bibinfo
  {journal} {Phys. Rev. B}\ }\textbf {\bibinfo {volume} {92}},\ \bibinfo
  {pages} {115119} (\bibinfo {year} {2015})}\BibitemShut {NoStop}%
\bibitem [{\citenamefont {Wu}\ \emph {et~al.}(2016)\citenamefont {Wu},
  \citenamefont {Qin}, \citenamefont {Liang}, \citenamefont {Fan},\ and\
  \citenamefont {Hu}}]{Wu2016iron}%
  \BibitemOpen
  \bibfield  {author} {\bibinfo {author} {\bibfnamefont {Xianxin}\ \bibnamefont
  {Wu}}, \bibinfo {author} {\bibfnamefont {Shengshan}\ \bibnamefont {Qin}},
  \bibinfo {author} {\bibfnamefont {Yi}~\bibnamefont {Liang}}, \bibinfo
  {author} {\bibfnamefont {Heng}\ \bibnamefont {Fan}}, \ and\ \bibinfo {author}
  {\bibfnamefont {Jiangping}\ \bibnamefont {Hu}},\ }\bibfield  {title}
  {\enquote {\bibinfo {title} {Topological characters in
  $\mathrm{Fe}({\mathrm{te}}_{1\ensuremath{-}x}{\mathrm{se}}_{x})$ thin
  films},}\ }\href {\doibase 10.1103/PhysRevB.93.115129} {\bibfield  {journal}
  {\bibinfo  {journal} {Phys. Rev. B}\ }\textbf {\bibinfo {volume} {93}},\
  \bibinfo {pages} {115129} (\bibinfo {year} {2016})}\BibitemShut {NoStop}%
\bibitem [{\citenamefont {Xu}\ \emph {et~al.}(2016)\citenamefont {Xu},
  \citenamefont {Lian}, \citenamefont {Tang}, \citenamefont {Qi},\ and\
  \citenamefont {Zhang}}]{Xu2016FeSC}%
  \BibitemOpen
  \bibfield  {author} {\bibinfo {author} {\bibfnamefont {Gang}\ \bibnamefont
  {Xu}}, \bibinfo {author} {\bibfnamefont {Biao}\ \bibnamefont {Lian}},
  \bibinfo {author} {\bibfnamefont {Peizhe}\ \bibnamefont {Tang}}, \bibinfo
  {author} {\bibfnamefont {Xiao-Liang}\ \bibnamefont {Qi}}, \ and\ \bibinfo
  {author} {\bibfnamefont {Shou-Cheng}\ \bibnamefont {Zhang}},\ }\bibfield
  {title} {\enquote {\bibinfo {title} {Topological superconductivity on the
  surface of fe-based superconductors},}\ }\href {\doibase
  10.1103/PhysRevLett.117.047001} {\bibfield  {journal} {\bibinfo  {journal}
  {Phys. Rev. Lett.}\ }\textbf {\bibinfo {volume} {117}},\ \bibinfo {pages}
  {047001} (\bibinfo {year} {2016})}\BibitemShut {NoStop}%
\bibitem [{\citenamefont {Zhang}\ \emph {et~al.}(2018)\citenamefont {Zhang},
  \citenamefont {Yaji}, \citenamefont {Hashimoto}, \citenamefont {Ota},
  \citenamefont {Kondo}, \citenamefont {Okazaki}, \citenamefont {Wang},
  \citenamefont {Wen}, \citenamefont {Gu}, \citenamefont {Ding} \emph
  {et~al.}}]{zhang2018iron}%
  \BibitemOpen
  \bibfield  {author} {\bibinfo {author} {\bibfnamefont {Peng}\ \bibnamefont
  {Zhang}}, \bibinfo {author} {\bibfnamefont {Koichiro}\ \bibnamefont {Yaji}},
  \bibinfo {author} {\bibfnamefont {Takahiro}\ \bibnamefont {Hashimoto}},
  \bibinfo {author} {\bibfnamefont {Yuichi}\ \bibnamefont {Ota}}, \bibinfo
  {author} {\bibfnamefont {Takeshi}\ \bibnamefont {Kondo}}, \bibinfo {author}
  {\bibfnamefont {Kozo}\ \bibnamefont {Okazaki}}, \bibinfo {author}
  {\bibfnamefont {Zhijun}\ \bibnamefont {Wang}}, \bibinfo {author}
  {\bibfnamefont {Jinsheng}\ \bibnamefont {Wen}}, \bibinfo {author}
  {\bibfnamefont {GD}~\bibnamefont {Gu}}, \bibinfo {author} {\bibfnamefont
  {Hong}\ \bibnamefont {Ding}},  \emph {et~al.},\ }\bibfield  {title} {\enquote
  {\bibinfo {title} {Observation of topological superconductivity on the
  surface of an iron-based superconductor},}\ }\href {\doibase
  10.1126/science.aan4596} {\bibfield  {journal} {\bibinfo  {journal}
  {Science}\ }\textbf {\bibinfo {volume} {360}},\ \bibinfo {pages} {182--186}
  (\bibinfo {year} {2018})}\BibitemShut {NoStop}%
\bibitem [{\citenamefont {Wang}\ \emph
  {et~al.}(2018{\natexlab{c}})\citenamefont {Wang}, \citenamefont {Kong},
  \citenamefont {Fan}, \citenamefont {Chen}, \citenamefont {Zhu}, \citenamefont
  {Liu}, \citenamefont {Cao}, \citenamefont {Sun}, \citenamefont {Du},
  \citenamefont {Schneeloch} \emph {et~al.}}]{wang2018evidence}%
  \BibitemOpen
  \bibfield  {author} {\bibinfo {author} {\bibfnamefont {Dongfei}\ \bibnamefont
  {Wang}}, \bibinfo {author} {\bibfnamefont {Lingyuan}\ \bibnamefont {Kong}},
  \bibinfo {author} {\bibfnamefont {Peng}\ \bibnamefont {Fan}}, \bibinfo
  {author} {\bibfnamefont {Hui}\ \bibnamefont {Chen}}, \bibinfo {author}
  {\bibfnamefont {Shiyu}\ \bibnamefont {Zhu}}, \bibinfo {author} {\bibfnamefont
  {Wenyao}\ \bibnamefont {Liu}}, \bibinfo {author} {\bibfnamefont
  {Lu}~\bibnamefont {Cao}}, \bibinfo {author} {\bibfnamefont {Yujie}\
  \bibnamefont {Sun}}, \bibinfo {author} {\bibfnamefont {Shixuan}\ \bibnamefont
  {Du}}, \bibinfo {author} {\bibfnamefont {John}\ \bibnamefont {Schneeloch}},
  \emph {et~al.},\ }\bibfield  {title} {\enquote {\bibinfo {title} {Evidence
  for majorana bound states in an iron-based superconductor},}\ }\href
  {\doibase 10.1126/science.aao1797} {\bibfield  {journal} {\bibinfo  {journal}
  {Science}\ }\textbf {\bibinfo {volume} {362}},\ \bibinfo {pages} {333--335}
  (\bibinfo {year} {2018}{\natexlab{c}})}\BibitemShut {NoStop}%
\bibitem [{\citenamefont {Kong}\ \emph {et~al.}(2019)\citenamefont {Kong},
  \citenamefont {Zhu}, \citenamefont {Papaj}, \citenamefont {Chen},
  \citenamefont {Cao}, \citenamefont {Isobe}, \citenamefont {Xing},
  \citenamefont {Liu}, \citenamefont {Wang}, \citenamefont {Fan} \emph
  {et~al.}}]{kong2019half}%
  \BibitemOpen
  \bibfield  {author} {\bibinfo {author} {\bibfnamefont {Lingyuan}\
  \bibnamefont {Kong}}, \bibinfo {author} {\bibfnamefont {Shiyu}\ \bibnamefont
  {Zhu}}, \bibinfo {author} {\bibfnamefont {Micha{\l}}\ \bibnamefont {Papaj}},
  \bibinfo {author} {\bibfnamefont {Hui}\ \bibnamefont {Chen}}, \bibinfo
  {author} {\bibfnamefont {Lu}~\bibnamefont {Cao}}, \bibinfo {author}
  {\bibfnamefont {Hiroki}\ \bibnamefont {Isobe}}, \bibinfo {author}
  {\bibfnamefont {Yuqing}\ \bibnamefont {Xing}}, \bibinfo {author}
  {\bibfnamefont {Wenyao}\ \bibnamefont {Liu}}, \bibinfo {author}
  {\bibfnamefont {Dongfei}\ \bibnamefont {Wang}}, \bibinfo {author}
  {\bibfnamefont {Peng}\ \bibnamefont {Fan}},  \emph {et~al.},\ }\bibfield
  {title} {\enquote {\bibinfo {title} {Half-integer level shift of vortex bound
  states in an iron-based superconductor},}\ }\href {\doibase
  10.1038/s41567-019-0630-5} {\bibfield  {journal} {\bibinfo  {journal} {Nature
  Physics}\ ,\ \bibinfo {pages} {1--7}} (\bibinfo {year} {2019})}\BibitemShut
  {NoStop}%
\bibitem [{\citenamefont {Machida}\ \emph {et~al.}(2019)\citenamefont
  {Machida}, \citenamefont {Sun}, \citenamefont {Pyon}, \citenamefont {Takeda},
  \citenamefont {Kohsaka}, \citenamefont {Hanaguri}, \citenamefont {Sasagawa},\
  and\ \citenamefont {Tamegai}}]{machida2019zero}%
  \BibitemOpen
  \bibfield  {author} {\bibinfo {author} {\bibfnamefont {T}~\bibnamefont
  {Machida}}, \bibinfo {author} {\bibfnamefont {Y}~\bibnamefont {Sun}},
  \bibinfo {author} {\bibfnamefont {S}~\bibnamefont {Pyon}}, \bibinfo {author}
  {\bibfnamefont {S}~\bibnamefont {Takeda}}, \bibinfo {author} {\bibfnamefont
  {Y}~\bibnamefont {Kohsaka}}, \bibinfo {author} {\bibfnamefont
  {T}~\bibnamefont {Hanaguri}}, \bibinfo {author} {\bibfnamefont
  {T}~\bibnamefont {Sasagawa}}, \ and\ \bibinfo {author} {\bibfnamefont
  {T}~\bibnamefont {Tamegai}},\ }\bibfield  {title} {\enquote {\bibinfo {title}
  {Zero-energy vortex bound state in the superconducting topological surface
  state of fe (se, te)},}\ }\href {\doibase 10.1038/s41563-019-0397-1}
  {\bibfield  {journal} {\bibinfo  {journal} {Nature materials}\ ,\ \bibinfo
  {pages} {1}} (\bibinfo {year} {2019})}\BibitemShut {NoStop}%
\bibitem [{\citenamefont {Liu}\ \emph {et~al.}(2018{\natexlab{b}})\citenamefont
  {Liu}, \citenamefont {Chen}, \citenamefont {Zhang}, \citenamefont {Peng},
  \citenamefont {Yan}, \citenamefont {Wen}, \citenamefont {Lou}, \citenamefont
  {Huang}, \citenamefont {Tian}, \citenamefont {Dong}, \citenamefont {Wang},
  \citenamefont {Bao}, \citenamefont {Wang}, \citenamefont {Yin}, \citenamefont
  {Zhao},\ and\ \citenamefont {Feng}}]{Liu2018MZM}%
  \BibitemOpen
  \bibfield  {author} {\bibinfo {author} {\bibfnamefont {Qin}\ \bibnamefont
  {Liu}}, \bibinfo {author} {\bibfnamefont {Chen}\ \bibnamefont {Chen}},
  \bibinfo {author} {\bibfnamefont {Tong}\ \bibnamefont {Zhang}}, \bibinfo
  {author} {\bibfnamefont {Rui}\ \bibnamefont {Peng}}, \bibinfo {author}
  {\bibfnamefont {Ya-Jun}\ \bibnamefont {Yan}}, \bibinfo {author}
  {\bibfnamefont {Chen-Hao-Ping}\ \bibnamefont {Wen}}, \bibinfo {author}
  {\bibfnamefont {Xia}\ \bibnamefont {Lou}}, \bibinfo {author} {\bibfnamefont
  {Yu-Long}\ \bibnamefont {Huang}}, \bibinfo {author} {\bibfnamefont
  {Jin-Peng}\ \bibnamefont {Tian}}, \bibinfo {author} {\bibfnamefont {Xiao-Li}\
  \bibnamefont {Dong}}, \bibinfo {author} {\bibfnamefont {Guang-Wei}\
  \bibnamefont {Wang}}, \bibinfo {author} {\bibfnamefont {Wei-Cheng}\
  \bibnamefont {Bao}}, \bibinfo {author} {\bibfnamefont {Qiang-Hua}\
  \bibnamefont {Wang}}, \bibinfo {author} {\bibfnamefont {Zhi-Ping}\
  \bibnamefont {Yin}}, \bibinfo {author} {\bibfnamefont {Zhong-Xian}\
  \bibnamefont {Zhao}}, \ and\ \bibinfo {author} {\bibfnamefont {Dong-Lai}\
  \bibnamefont {Feng}},\ }\bibfield  {title} {\enquote {\bibinfo {title}
  {Robust and clean majorana zero mode in the vortex core of high-temperature
  superconductor
  $\mathbf{(}{\mathrm{li}}_{0.84}{\mathrm{fe}}_{0.16}\mathbf{)}\mathrm{OHFeSe}$},}\
  }\href {\doibase 10.1103/PhysRevX.8.041056} {\bibfield  {journal} {\bibinfo
  {journal} {Phys. Rev. X}\ }\textbf {\bibinfo {volume} {8}},\ \bibinfo {pages}
  {041056} (\bibinfo {year} {2018}{\natexlab{b}})}\BibitemShut {NoStop}%
\bibitem [{\citenamefont {Chen}\ \emph {et~al.}(2019)\citenamefont {Chen},
  \citenamefont {Liu}, \citenamefont {Zhang}, \citenamefont {Li}, \citenamefont
  {Shen}, \citenamefont {Dong}, \citenamefont {Zhao}, \citenamefont {Zhang},\
  and\ \citenamefont {Feng}}]{chen2019quantized}%
  \BibitemOpen
  \bibfield  {author} {\bibinfo {author} {\bibfnamefont {C}~\bibnamefont
  {Chen}}, \bibinfo {author} {\bibfnamefont {Q}~\bibnamefont {Liu}}, \bibinfo
  {author} {\bibfnamefont {TZ}~\bibnamefont {Zhang}}, \bibinfo {author}
  {\bibfnamefont {D}~\bibnamefont {Li}}, \bibinfo {author} {\bibfnamefont
  {PP}~\bibnamefont {Shen}}, \bibinfo {author} {\bibfnamefont {XL}~\bibnamefont
  {Dong}}, \bibinfo {author} {\bibfnamefont {Z-X}\ \bibnamefont {Zhao}},
  \bibinfo {author} {\bibfnamefont {T}~\bibnamefont {Zhang}}, \ and\ \bibinfo
  {author} {\bibfnamefont {DL}~\bibnamefont {Feng}},\ }\bibfield  {title}
  {\enquote {\bibinfo {title} {Quantized conductance of majorana zero mode in
  the vortex of the topological superconductor (li0. 84fe0. 16) ohfese},}\
  }\href {\doibase 10.1088/0256-307x/36/5/057403} {\bibfield  {journal}
  {\bibinfo  {journal} {Chinese Physics Letters}\ }\textbf {\bibinfo {volume}
  {36}},\ \bibinfo {pages} {057403} (\bibinfo {year} {2019})}\BibitemShut
  {NoStop}%
\bibitem [{\citenamefont {Zhu}\ \emph {et~al.}(2020)\citenamefont {Zhu},
  \citenamefont {Kong}, \citenamefont {Cao}, \citenamefont {Chen},
  \citenamefont {Papaj}, \citenamefont {Du}, \citenamefont {Xing},
  \citenamefont {Liu}, \citenamefont {Wang}, \citenamefont {Shen},
  \citenamefont {Yang}, \citenamefont {Schneeloch}, \citenamefont {Zhong},
  \citenamefont {Gu}, \citenamefont {Fu}, \citenamefont {Zhang}, \citenamefont
  {Ding},\ and\ \citenamefont {Gao}}]{Zhu2019MZM}%
  \BibitemOpen
  \bibfield  {author} {\bibinfo {author} {\bibfnamefont {Shiyu}\ \bibnamefont
  {Zhu}}, \bibinfo {author} {\bibfnamefont {Lingyuan}\ \bibnamefont {Kong}},
  \bibinfo {author} {\bibfnamefont {Lu}~\bibnamefont {Cao}}, \bibinfo {author}
  {\bibfnamefont {Hui}\ \bibnamefont {Chen}}, \bibinfo {author} {\bibfnamefont
  {Micha{\l}}\ \bibnamefont {Papaj}}, \bibinfo {author} {\bibfnamefont
  {Shixuan}\ \bibnamefont {Du}}, \bibinfo {author} {\bibfnamefont {Yuqing}\
  \bibnamefont {Xing}}, \bibinfo {author} {\bibfnamefont {Wenyao}\ \bibnamefont
  {Liu}}, \bibinfo {author} {\bibfnamefont {Dongfei}\ \bibnamefont {Wang}},
  \bibinfo {author} {\bibfnamefont {Chengmin}\ \bibnamefont {Shen}}, \bibinfo
  {author} {\bibfnamefont {Fazhi}\ \bibnamefont {Yang}}, \bibinfo {author}
  {\bibfnamefont {John}\ \bibnamefont {Schneeloch}}, \bibinfo {author}
  {\bibfnamefont {Ruidan}\ \bibnamefont {Zhong}}, \bibinfo {author}
  {\bibfnamefont {Genda}\ \bibnamefont {Gu}}, \bibinfo {author} {\bibfnamefont
  {Liang}\ \bibnamefont {Fu}}, \bibinfo {author} {\bibfnamefont {Yu-Yang}\
  \bibnamefont {Zhang}}, \bibinfo {author} {\bibfnamefont {Hong}\ \bibnamefont
  {Ding}}, \ and\ \bibinfo {author} {\bibfnamefont {Hong-Jun}\ \bibnamefont
  {Gao}},\ }\bibfield  {title} {\enquote {\bibinfo {title} {Nearly quantized
  conductance plateau of vortex zero mode in an iron-based superconductor},}\
  }\href {\doibase 10.1126/science.aax0274} {\bibfield  {journal} {\bibinfo
  {journal} {Science}\ }\textbf {\bibinfo {volume} {367}},\ \bibinfo {pages}
  {189--192} (\bibinfo {year} {2020})}\BibitemShut {NoStop}%
\bibitem [{\citenamefont {Mazin}\ \emph {et~al.}(2008)\citenamefont {Mazin},
  \citenamefont {Singh}, \citenamefont {Johannes},\ and\ \citenamefont
  {Du}}]{Mazin2008}%
  \BibitemOpen
  \bibfield  {author} {\bibinfo {author} {\bibfnamefont {I.~I.}\ \bibnamefont
  {Mazin}}, \bibinfo {author} {\bibfnamefont {D.~J.}\ \bibnamefont {Singh}},
  \bibinfo {author} {\bibfnamefont {M.~D.}\ \bibnamefont {Johannes}}, \ and\
  \bibinfo {author} {\bibfnamefont {M.~H.}\ \bibnamefont {Du}},\ }\bibfield
  {title} {\enquote {\bibinfo {title} {Unconventional superconductivity with a
  sign reversal in the order parameter of
  ${\mathrm{lafeaso}}_{1\ensuremath{-}x}{\mathrm{f}}_{x}$},}\ }\href {\doibase
  10.1103/PhysRevLett.101.057003} {\bibfield  {journal} {\bibinfo  {journal}
  {Phys. Rev. Lett.}\ }\textbf {\bibinfo {volume} {101}},\ \bibinfo {pages}
  {057003} (\bibinfo {year} {2008})}\BibitemShut {NoStop}%
\bibitem [{\citenamefont {Wang}\ and\ \citenamefont
  {Lee}(2011)}]{Wang2011iron}%
  \BibitemOpen
  \bibfield  {author} {\bibinfo {author} {\bibfnamefont {Fa}~\bibnamefont
  {Wang}}\ and\ \bibinfo {author} {\bibfnamefont {Dung-Hai}\ \bibnamefont
  {Lee}},\ }\bibfield  {title} {\enquote {\bibinfo {title} {The
  electron-pairing mechanism of iron-based superconductors},}\ }\href {\doibase
  10.1126/science.1200182} {\bibfield  {journal} {\bibinfo  {journal}
  {Science}\ }\textbf {\bibinfo {volume} {332}},\ \bibinfo {pages} {200--204}
  (\bibinfo {year} {2011})}\BibitemShut {NoStop}%
\bibitem [{\citenamefont {Schnyder}\ \emph {et~al.}(2008)\citenamefont
  {Schnyder}, \citenamefont {Ryu}, \citenamefont {Furusaki},\ and\
  \citenamefont {Ludwig}}]{schnyder2008}%
  \BibitemOpen
  \bibfield  {author} {\bibinfo {author} {\bibfnamefont {Andreas~P.}\
  \bibnamefont {Schnyder}}, \bibinfo {author} {\bibfnamefont {Shinsei}\
  \bibnamefont {Ryu}}, \bibinfo {author} {\bibfnamefont {Akira}\ \bibnamefont
  {Furusaki}}, \ and\ \bibinfo {author} {\bibfnamefont {Andreas W.~W.}\
  \bibnamefont {Ludwig}},\ }\bibfield  {title} {\enquote {\bibinfo {title}
  {Classification of topological insulators and superconductors in three
  spatial dimensions},}\ }\href {\doibase 10.1103/PhysRevB.78.195125}
  {\bibfield  {journal} {\bibinfo  {journal} {Phys. Rev. B}\ }\textbf {\bibinfo
  {volume} {78}},\ \bibinfo {pages} {195125} (\bibinfo {year}
  {2008})}\BibitemShut {NoStop}%
\bibitem [{\citenamefont {Kitaev}(2009)}]{kitaev2009}%
  \BibitemOpen
  \bibfield  {author} {\bibinfo {author} {\bibfnamefont {Alexei}\ \bibnamefont
  {Kitaev}},\ }\bibfield  {title} {\enquote {\bibinfo {title} {Periodic table
  for topological insulators and superconductors},}\ }in\ \href {\doibase
  10.1063/1.3149495} {\emph {\bibinfo {booktitle} {AIP Conference
  Proceedings}}},\ Vol.\ \bibinfo {volume} {1134}\ (\bibinfo {organization}
  {AIP},\ \bibinfo {year} {2009})\ pp.\ \bibinfo {pages} {22--30}\BibitemShut
  {NoStop}%
\bibitem [{\citenamefont {Haim}\ and\ \citenamefont
  {Oreg}(2019)}]{Haim2019review}%
  \BibitemOpen
  \bibfield  {author} {\bibinfo {author} {\bibfnamefont {Arbel}\ \bibnamefont
  {Haim}}\ and\ \bibinfo {author} {\bibfnamefont {Yuval}\ \bibnamefont
  {Oreg}},\ }\bibfield  {title} {\enquote {\bibinfo {title}
  {Time-reversal-invariant topological superconductivity in one and two
  dimensions},}\ }\href {\doibase
  https://doi.org/10.1016/j.physrep.2019.08.002} {\bibfield  {journal}
  {\bibinfo  {journal} {Physics Reports}\ }\textbf {\bibinfo {volume} {825}},\
  \bibinfo {pages} {1 -- 48} (\bibinfo {year} {2019})},\ \bibinfo {note}
  {time-reversal-invariant topological superconductivity in one and two
  dimensions}\BibitemShut {NoStop}%
\bibitem [{\citenamefont {Qi}\ \emph {et~al.}(2010)\citenamefont {Qi},
  \citenamefont {Hughes},\ and\ \citenamefont {Zhang}}]{qi2010d}%
  \BibitemOpen
  \bibfield  {author} {\bibinfo {author} {\bibfnamefont {Xiao-Liang}\
  \bibnamefont {Qi}}, \bibinfo {author} {\bibfnamefont {Taylor~L.}\
  \bibnamefont {Hughes}}, \ and\ \bibinfo {author} {\bibfnamefont {Shou-Cheng}\
  \bibnamefont {Zhang}},\ }\bibfield  {title} {\enquote {\bibinfo {title}
  {Topological invariants for the fermi surface of a time-reversal-invariant
  superconductor},}\ }\href {\doibase 10.1103/PhysRevB.81.134508} {\bibfield
  {journal} {\bibinfo  {journal} {Phys. Rev. B}\ }\textbf {\bibinfo {volume}
  {81}},\ \bibinfo {pages} {134508} (\bibinfo {year} {2010})}\BibitemShut
  {NoStop}%
\bibitem [{sup()}]{supplemental}%
  \BibitemOpen
  \href@noop {} {}\bibinfo {howpublished} {The supplemental material contains
  the derivation of boundary Hamiltonians.}\BibitemShut {Stop}%
\bibitem [{\citenamefont {Wong}\ and\ \citenamefont
  {Law}(2012)}]{Wong2012majorana}%
  \BibitemOpen
  \bibfield  {author} {\bibinfo {author} {\bibfnamefont {Chris L.~M.}\
  \bibnamefont {Wong}}\ and\ \bibinfo {author} {\bibfnamefont {K.~T.}\
  \bibnamefont {Law}},\ }\bibfield  {title} {\enquote {\bibinfo {title}
  {Majorana kramers doublets in ${d}_{{x}^{2}\ensuremath{-}{y}^{2}}$-wave
  superconductors with rashba spin-orbit coupling},}\ }\href {\doibase
  10.1103/PhysRevB.86.184516} {\bibfield  {journal} {\bibinfo  {journal} {Phys.
  Rev. B}\ }\textbf {\bibinfo {volume} {86}},\ \bibinfo {pages} {184516}
  (\bibinfo {year} {2012})}\BibitemShut {NoStop}%
\bibitem [{\citenamefont {Zhang}\ \emph {et~al.}(2013)\citenamefont {Zhang},
  \citenamefont {Kane},\ and\ \citenamefont {Mele}}]{zhang2013kramers}%
  \BibitemOpen
  \bibfield  {author} {\bibinfo {author} {\bibfnamefont {Fan}\ \bibnamefont
  {Zhang}}, \bibinfo {author} {\bibfnamefont {C.~L.}\ \bibnamefont {Kane}}, \
  and\ \bibinfo {author} {\bibfnamefont {E.~J.}\ \bibnamefont {Mele}},\
  }\bibfield  {title} {\enquote {\bibinfo {title} {Time-reversal-invariant
  topological superconductivity and majorana kramers pairs},}\ }\href {\doibase
  10.1103/PhysRevLett.111.056402} {\bibfield  {journal} {\bibinfo  {journal}
  {Phys. Rev. Lett.}\ }\textbf {\bibinfo {volume} {111}},\ \bibinfo {pages}
  {056402} (\bibinfo {year} {2013})}\BibitemShut {NoStop}%
\bibitem [{\citenamefont {Keselman}\ \emph {et~al.}(2013)\citenamefont
  {Keselman}, \citenamefont {Fu}, \citenamefont {Stern},\ and\ \citenamefont
  {Berg}}]{keselman2013}%
  \BibitemOpen
  \bibfield  {author} {\bibinfo {author} {\bibfnamefont {Anna}\ \bibnamefont
  {Keselman}}, \bibinfo {author} {\bibfnamefont {Liang}\ \bibnamefont {Fu}},
  \bibinfo {author} {\bibfnamefont {Ady}\ \bibnamefont {Stern}}, \ and\
  \bibinfo {author} {\bibfnamefont {Erez}\ \bibnamefont {Berg}},\ }\bibfield
  {title} {\enquote {\bibinfo {title} {Inducing time-reversal-invariant
  topological superconductivity and fermion parity pumping in quantum wires},}\
  }\href {\doibase 10.1103/PhysRevLett.111.116402} {\bibfield  {journal}
  {\bibinfo  {journal} {Phys. Rev. Lett.}\ }\textbf {\bibinfo {volume} {111}},\
  \bibinfo {pages} {116402} (\bibinfo {year} {2013})}\BibitemShut {NoStop}%
\bibitem [{\citenamefont {Haim}\ \emph {et~al.}(2014)\citenamefont {Haim},
  \citenamefont {Keselman}, \citenamefont {Berg},\ and\ \citenamefont
  {Oreg}}]{haim2014}%
  \BibitemOpen
  \bibfield  {author} {\bibinfo {author} {\bibfnamefont {Arbel}\ \bibnamefont
  {Haim}}, \bibinfo {author} {\bibfnamefont {Anna}\ \bibnamefont {Keselman}},
  \bibinfo {author} {\bibfnamefont {Erez}\ \bibnamefont {Berg}}, \ and\
  \bibinfo {author} {\bibfnamefont {Yuval}\ \bibnamefont {Oreg}},\ }\bibfield
  {title} {\enquote {\bibinfo {title} {Time-reversal-invariant topological
  superconductivity induced by repulsive interactions in quantum wires},}\
  }\href {\doibase 10.1103/PhysRevB.89.220504} {\bibfield  {journal} {\bibinfo
  {journal} {Phys. Rev. B}\ }\textbf {\bibinfo {volume} {89}},\ \bibinfo
  {pages} {220504} (\bibinfo {year} {2014})}\BibitemShut {NoStop}%
\bibitem [{\citenamefont {Gaidamauskas}\ \emph {et~al.}(2014)\citenamefont
  {Gaidamauskas}, \citenamefont {Paaske},\ and\ \citenamefont
  {Flensberg}}]{Gaidamauskas2014}%
  \BibitemOpen
  \bibfield  {author} {\bibinfo {author} {\bibfnamefont {Erikas}\ \bibnamefont
  {Gaidamauskas}}, \bibinfo {author} {\bibfnamefont {Jens}\ \bibnamefont
  {Paaske}}, \ and\ \bibinfo {author} {\bibfnamefont {Karsten}\ \bibnamefont
  {Flensberg}},\ }\bibfield  {title} {\enquote {\bibinfo {title} {Majorana
  bound states in two-channel time-reversal-symmetric nanowire systems},}\
  }\href {\doibase 10.1103/PhysRevLett.112.126402} {\bibfield  {journal}
  {\bibinfo  {journal} {Phys. Rev. Lett.}\ }\textbf {\bibinfo {volume} {112}},\
  \bibinfo {pages} {126402} (\bibinfo {year} {2014})}\BibitemShut {NoStop}%
\bibitem [{\citenamefont {Schrade}\ \emph {et~al.}(2015)\citenamefont
  {Schrade}, \citenamefont {Zyuzin}, \citenamefont {Klinovaja},\ and\
  \citenamefont {Loss}}]{Schrade2015}%
  \BibitemOpen
  \bibfield  {author} {\bibinfo {author} {\bibfnamefont {Constantin}\
  \bibnamefont {Schrade}}, \bibinfo {author} {\bibfnamefont {A.~A.}\
  \bibnamefont {Zyuzin}}, \bibinfo {author} {\bibfnamefont {Jelena}\
  \bibnamefont {Klinovaja}}, \ and\ \bibinfo {author} {\bibfnamefont {Daniel}\
  \bibnamefont {Loss}},\ }\bibfield  {title} {\enquote {\bibinfo {title}
  {Proximity-induced $\ensuremath{\pi}$ josephson junctions in topological
  insulators and kramers pairs of majorana fermions},}\ }\href {\doibase
  10.1103/PhysRevLett.115.237001} {\bibfield  {journal} {\bibinfo  {journal}
  {Phys. Rev. Lett.}\ }\textbf {\bibinfo {volume} {115}},\ \bibinfo {pages}
  {237001} (\bibinfo {year} {2015})}\BibitemShut {NoStop}%
\bibitem [{\citenamefont {Deng}\ \emph {et~al.}(2012)\citenamefont {Deng},
  \citenamefont {Viola},\ and\ \citenamefont {Ortiz}}]{Deng2012TRITSC}%
  \BibitemOpen
  \bibfield  {author} {\bibinfo {author} {\bibfnamefont {Shusa}\ \bibnamefont
  {Deng}}, \bibinfo {author} {\bibfnamefont {Lorenza}\ \bibnamefont {Viola}}, \
  and\ \bibinfo {author} {\bibfnamefont {Gerardo}\ \bibnamefont {Ortiz}},\
  }\bibfield  {title} {\enquote {\bibinfo {title} {Majorana modes in
  time-reversal invariant $s$-wave topological superconductors},}\ }\href
  {\doibase 10.1103/PhysRevLett.108.036803} {\bibfield  {journal} {\bibinfo
  {journal} {Phys. Rev. Lett.}\ }\textbf {\bibinfo {volume} {108}},\ \bibinfo
  {pages} {036803} (\bibinfo {year} {2012})}\BibitemShut {NoStop}%
\bibitem [{\citenamefont {Kheirkhah}\ \emph
  {et~al.}(2020{\natexlab{c}})\citenamefont {Kheirkhah}, \citenamefont {Yan},\
  and\ \citenamefont {Marsiglio}}]{kheirkhah2020vortex}%
  \BibitemOpen
  \bibfield  {author} {\bibinfo {author} {\bibfnamefont {Majid}\ \bibnamefont
  {Kheirkhah}}, \bibinfo {author} {\bibfnamefont {Zhongbo}\ \bibnamefont
  {Yan}}, \ and\ \bibinfo {author} {\bibfnamefont {Frank}\ \bibnamefont
  {Marsiglio}},\ }\href@noop {} {\enquote {\bibinfo {title} {Vortex line
  topology in iron-based superconductors with and without second-order
  topology},}\ } (\bibinfo {year} {2020}{\natexlab{c}}),\ \Eprint
  {http://arxiv.org/abs/2007.10326} {arXiv:2007.10326 [cond-mat.supr-con]}
  \BibitemShut {NoStop}%
\end{thebibliography}%

\begin{widetext}
\clearpage
\begin{center}
\textbf{\large Supplemental Material ``Boundary Topological Superconductors''}\\
\vspace{4mm}
{Bo-Xuan Li$^{1}$, Zhongbo Yan$^{1,*}$}\\
\vspace{2mm}
{\em $^1$School of Physics, Sun Yat-sen University, Guangzhou, 510275, China}
\end{center}

\setcounter{equation}{0}
\setcounter{figure}{0}
\setcounter{table}{0}
\makeatletter
\renewcommand{\theequation}{S\arabic{equation}}
\renewcommand{\thefigure}{S\arabic{figure}}
\renewcommand{\bibnumfmt}[1]{[S#1]}

This supplemental material contains the derivation of boundary Hamiltonians. We start with
the normal-state Hamiltonian in two dimensions (2D), which is given by
\begin{eqnarray}
H_{0}(\bk)&=&(\epsilon_{1}\cos k_{1}+\epsilon_{2}\cos k_{2})\Sigma_{00}+(m_{0}-m_{1}\cos k_{1}-m_{2}\cos k_{2})\Sigma_{03}\nonumber\\
&&+\lambda_{1}\sin k_{1}\Sigma_{31}+\lambda_{2}\sin k_{2}\Sigma_{02},
\end{eqnarray}
where $\Sigma_{ij}=s_{i}\otimes \sigma_{j}$, with
the Pauli matrices $s_{i}$ and $\sigma_{j}$ acting respectively
on the spin $(\uparrow,\downarrow)$ and orbital $(a,b)$ degrees of freedom. $s_{0}$ and $\sigma_{0}$ are
two-by-two unit matrices.

Consider that the band inversion occurs at the two time-reversal invariant (TRI) momenta $(k_{1},k_{2})=(0,0)$ and $(\pi,0)$,
we first make a Taylor expansion around $k_{2}=0$ to the second order of $k_{2}$. Accordingly, we have
\begin{eqnarray}
H_{0}(k_{1},k_{2})=[\epsilon(k_{1})-\frac{\epsilon_{2}}{2}k_{2}^{2}]\Sigma_{00}+[m(k_{1})+\frac{m_{2}}{2}k_{2}^{2}]\Sigma_{03}
+\lambda_{1}\sin k_{1}\Sigma_{31}+\lambda_{2}k_{2}\Sigma_{02},
\end{eqnarray}
where $\epsilon(k_{1})=\epsilon_{2}+\epsilon_{1}\cos k_{1}$ and $m(k_{1})=m_{0}-m_{2}-m_{1}\cos k_{1}$.
In the following, we consider $m_{2}$ and $\lambda_{2}$ to be positive. According to the band inversion,
we have $m(k_{1})<0$ for an arbitrary $k_{1}\in(-\pi,\pi)$.

To obtain the gapless states on the $x_{2}$-normal edges, we further consider a half-infinity sample which occupies
the region $0\leq x_{2}\leq +\infty$. Because the translational symmetry is broken in the $x_{2}$ direction,
the wave vector $k_{2}$ needs to be replaced by $-i\partial_{x_{2}}$. Accordingly, the Hamiltonian becomes
\begin{eqnarray}
H_{0}(k_{1},-i\partial_{x_{2}})&=&[\epsilon(k_{1})+\frac{\epsilon_{2}}{2}\partial_{x_{2}}^{2}]\Sigma_{00}
+[m(k_{1})-\frac{m_{2}}{2}\partial_{x_{2}}^{2}]\Sigma_{03}\nonumber\\
&&+\lambda_{1}\sin k_{1}\Sigma_{31}-i\lambda_{2}\partial_{x_{2}}\Sigma_{02}.
\end{eqnarray}
Next, we divide the Hamiltonian into two parts, i.e., $H_{0}=H_{1}+H_{2}$, where
\begin{eqnarray}
H_{1}(k_{1},-i\partial_{x_{2}})&=&[m(k_{1})-\frac{m_{2}}{2}\partial_{x_{2}}^{2}]\Sigma_{03}-i\lambda_{2}\partial_{x_{2}}\Sigma_{02},\nonumber\\
H_{2}(k_{1},-i\partial_{x_{2}})&=&[\epsilon(k_{1})+\frac{\epsilon_{2}}{2}\partial_{x_{2}}^{2}]\Sigma_{00}+\lambda_{1}\sin k_{1}\Sigma_{31}.
\end{eqnarray}
We first solve the eigenvalue equation $H_{1}(k_{1},-i\partial_{x_{2}})\psi(x_{2})=E\psi(x_{2})$ under
the boundary condition $\psi(0)=\psi(+\infty)=0$. It is readily found that there are two zero-energy solutions
for an arbitrary $k_{1}\in(-\pi,\pi)$. The two solutions take the form
\begin{eqnarray}
\psi_{\alpha=1,2}(x_{2})=\mathcal{N}\sin(\kappa_{1}x_{2})e^{-\kappa_{2}x_{2}}e^{ik_{1}x_{1}}\chi_{\alpha},
\end{eqnarray}
with normalization given by $|\mathcal{N}|^{2}=4|\kappa_{2}(\kappa_{1}^{2}+\kappa_{2}^{2})/\kappa_{1}^{2}|$,
where $\kappa_{1}=\sqrt{-\frac{2m(k_{1})}{m_{2}}-\frac{\lambda_{2}^{2}}{m_{2}^{2}}}$, and $\kappa_{2}=\frac{\lambda_{2}}{m_{2}}$.
Furthermore, $\chi_{\alpha}$ satisfy $\Sigma_{01}\chi_{\alpha}=\chi_{\alpha}$, which can be chosen as
\begin{eqnarray}
\chi_{1}&=&|s_{3}=1, \sigma_{1}=1\rangle, \nonumber\\
\chi_{2}&=&|s_{3}=-1, \sigma_{1}=1\rangle.
\end{eqnarray}
By projecting $H_{2}$ into the subspace expanded by $\chi_{1}$ and $\chi_{2}$, we obtain
the boundary Hamiltonian, which is given by
\begin{eqnarray}
[H_{e}(k_{1})]_{\alpha\beta}&=&\int_{0}^{+\infty}dx_{2}\psi_{\alpha}^{\dag}(x_{2})H_{2}(k_{1},-i\partial_{x_{2}})\psi_{\beta}(x_{2})\nonumber\\
&=&\left\{[\epsilon(k_{1})+\frac{m_{1}(k_{1})\epsilon_{2}}{m_{2}}]s_{0}+\lambda_{1}\sin k_{1}s_{3}\right\}_{\alpha\beta},
\end{eqnarray}
or equivalently,
\begin{eqnarray}
H_{e}(k_{1})&=&[\epsilon(k_{1})+\frac{m_{1}(k_{1})\epsilon_{2}}{m_{2}}]s_{0}+\lambda_{1}\sin k_{1}s_{3}\nonumber\\
&=&[\frac{m_{0}\epsilon_{2}}{m_{2}}+\epsilon_{1}\cos k_{1}-\frac{m_{1}\epsilon_{2}}{m_{2}}\cos k_{1}]s_{0}+\lambda_{1}\sin k_{1}s_{3}.\label{NH}
\end{eqnarray}
For the conduction-valence symmetric case, i.e., $\epsilon_{1,2}=0$, Eq.(\ref{NH}) reduces to Eq.(\ref{ef}) of
the main text, and the analytical results  perfectly agree with the numerical results, as shown in Fig.\ref{match}(a).
When conduction-valence asymmetry is present and strong, the analytical results still agree
with the numerical results well, as shown in Figs.\ref{match}(b)(c).
\begin{figure}
\subfigure{\includegraphics[width=15cm, height=4cm]{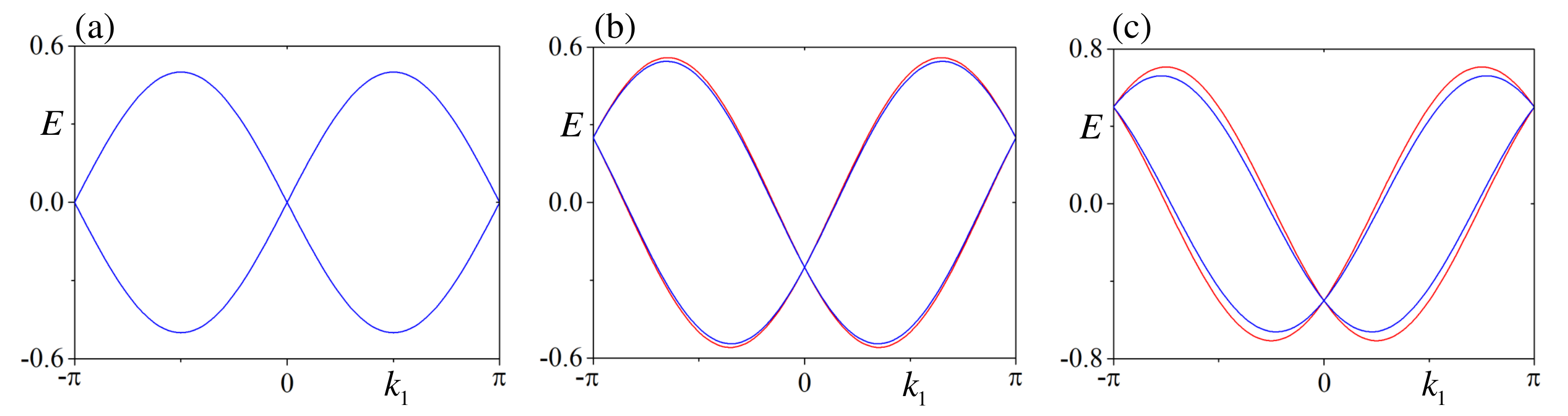}}
\caption{ Dispersions of the helical gapless states on the $x_{2}$-normal edges. The blue curves
correspond to numerical results, and the red curves correspond to analytical results. (a) $\epsilon_{1}=\epsilon_{2}=0$,
the numerical and analytical results perfectly agree with each other. (b) $\epsilon_{1}=0$, $\epsilon_{2}=0.5$,
the numerical and analytical results show small deviation away from time-reversal invariant momenta. (c) $\epsilon_{1}=0$, $\epsilon_{2}=1$.
The increase of conduction-valence asymmetry increases the deviation between numerical and analytical results, suggesting
that the expansion of $k_{2}$ to the second order is no longer sufficient.
}  \label{match}
\end{figure}

After taking into account the $s_{\pm}$-wave pairings, the Hamiltonian needs to be generalized
as
\begin{eqnarray}
H_{\rm BdG}(\bk)&=&[(\epsilon_{1}\cos k_{1}+\epsilon_{2}\cos k_{2})-\mu]\Sigma_{300}+(m_{0}-m_{1}\cos k_{1}-m_{2}\cos k_{2})\Sigma_{303}\nonumber\\
&&+\lambda_{1}\sin k_{1}\Sigma_{031}+\lambda_{2}\sin k_{2}\Sigma_{302}+[\Delta_{0}-\Delta_{1}(\cos k_{1}+\cos k_{2})]\Sigma_{220}.
\end{eqnarray}
where $\Sigma_{ijk}=\tau_{i}\otimes s_{j}\otimes \sigma_{k}$, with the new Pauli matrices $\tau_{i}$ acting
on the particle-hole degrees of freedom. Similarly, we make an expansion about $k_{2}$, and then do the replacement $k_{2}\rightarrow -i\partial_{x_{2}}$ and decomposition $H_{\rm BdG}=H_{1}+H_{2}$. Accordingly, we have
\begin{eqnarray}
H_{1}(k_{1},-i\partial_{x_{2}})&=&[m(k_{1})-\frac{m_{2}}{2}\partial_{x_{2}}^{2}]\Sigma_{303}-i\lambda_{2}\partial_{x_{2}}\Sigma_{302},\nonumber\\
H_{2}(k_{1},-i\partial_{x_{2}})&=&[\epsilon(k_{1})+\frac{\epsilon_{2}}{2}\partial_{x_{2}}^{2}-\mu]\Sigma_{300}+\lambda_{1}\sin k_{1}\Sigma_{031}
+[\Delta_{0}-\Delta_{1}\cos k_{1}-(\Delta_{1}+\frac{\Delta_{1}}{2}\partial_{x_{2}}^{2})]\Sigma_{220}.
\end{eqnarray}
Because of the increase of  particle-hole degrees of freedom, now there are four zero-energy solutions of the eigenvalue
equation $H_{1}(k_{1},-i\partial_{x_{2}})\psi_{\alpha}(x_{2})=E\psi_{\alpha}(x_{2})$. The general form of $\psi_{\alpha}(x_{2})$
still reads
\begin{eqnarray}
\psi_{\alpha=1,2,3,4}(x_{2})=\mathcal{N}\sin(\kappa_{1}x_{2})e^{-\kappa_{2}x_{2}}e^{ik_{1}x_{1}}\chi_{\alpha},
\end{eqnarray}
but now $\chi_{\alpha}$ satisfy $\Sigma_{001}\chi_{\alpha}=\chi_{\alpha}$, which can be chosen as
\begin{eqnarray}
\chi_{1}&=&|\tau_{3}=1, s_{3}=1, \sigma_{1}=1\rangle, \nonumber\\
\chi_{2}&=&|\tau_{3}=1, s_{3}=-1, \sigma_{1}=1\rangle, \nonumber\\
\chi_{3}&=&|\tau_{3}=-1, s_{3}=1, \sigma_{1}=1\rangle, \nonumber\\
\chi_{4}&=&|\tau_{3}=-1, s_{3}=-1, \sigma_{1}=1\rangle.
\end{eqnarray}
By projecting $H_{2}$ into the four-dimensional subspace expanded by $\chi_{1,2,3,4}$, we obtain
the boundary Hamiltonian, which is given by
\begin{eqnarray}
[H_{e}(k_{1})]_{\alpha\beta}&=&\int_{0}^{+\infty}dx_{2}\psi_{\alpha}^{\dag}(x_{2})H_{2}(k_{1},-i\partial_{x_{2}})\psi_{\beta}(x_{2})\nonumber\\
&=&\left\{[\epsilon(k_{1})+\frac{m_{1}(k_{1})\epsilon_{2}}{m_{2}}-\mu]\tau_{3}\otimes s_{0}+\lambda_{1}\sin k_{1}\tau_{0}\otimes s_{3}+[\Delta_{0}-\Delta_{1}\cos k_{1}-\Delta_{2}(k_{1})]\tau_{2}\otimes s_{2}\right\}_{\alpha\beta},
\end{eqnarray}
or equivalently,
\begin{eqnarray}
H_{e}(k_{1})
=\left(\frac{m_{0}\epsilon_{2}}{m_{2}}+\epsilon_{1}\cos k_{1}-\frac{m_{1}\epsilon_{2}}{m_{2}}\cos k_{1}-\mu\right)\tau_{3}\otimes s_{0}+\lambda_{1}\sin k_{1}\tau_{0}\otimes s_{3}+[\Delta_{0}-\Delta_{1}\cos k_{1}-\Delta_{2}(k_{1})]\tau_{2}\otimes s_{2},\quad\label{SH}
\end{eqnarray}
where, to the first two orders,
\begin{eqnarray}
\Delta_{2}(k_{1})&=&|\mathcal{N}^{2}|\int_{0}^{+\infty}dx_{2}\sin(\kappa_{1}x_{2})e^{-\kappa_{2}x_{2}}
[\Delta_{1}+\frac{\Delta_{1}}{2}\partial_{x_{2}}^{2}]\sin(\kappa_{1}x_{2})e^{-\kappa_{2}x_{2}}\nonumber\\
&=&\Delta_{1}+\frac{\Delta_{1}m(k_{1})}{m_{2}}\nonumber\\
&=&\frac{m_{0}-m_{1}\cos k_{1}}{m_{2}}.
\end{eqnarray}
Eq.(\ref{SH}) is just the Eq.(\ref{BH}) of the main text. According to Eq.(\ref{SH}), the energy bands are given by
\begin{eqnarray}
E_{\pm,\pm}(k_{1})=\pm\sqrt{\left(\frac{m_{0}\epsilon_{2}}{m_{2}}+\epsilon_{1}\cos k_{1}-\frac{m_{1}\epsilon_{2}}{m_{2}}\cos k_{1}-\mu\pm \lambda_{1}|\sin k_{1}|\right)^{2}+[\Delta_{0}-\Delta_{1}\cos k_{1}-\Delta_{2}(k_{1})]^{2}}.
\end{eqnarray}
For the conduction-valence symmetric case, the numerical and analytical results again show excellent agreement, as shown in Fig.\ref{match2}(a).

Similarly, we can derive the boundary Hamiltonian for the 3D case. Since the derivation is similar and straightforward, we neglect the process
and just show that the agreement of numerical and analytical results, as shown in  Fig.\ref{match2}(b).

\begin{figure}
\subfigure{\includegraphics[width=10cm, height=4cm]{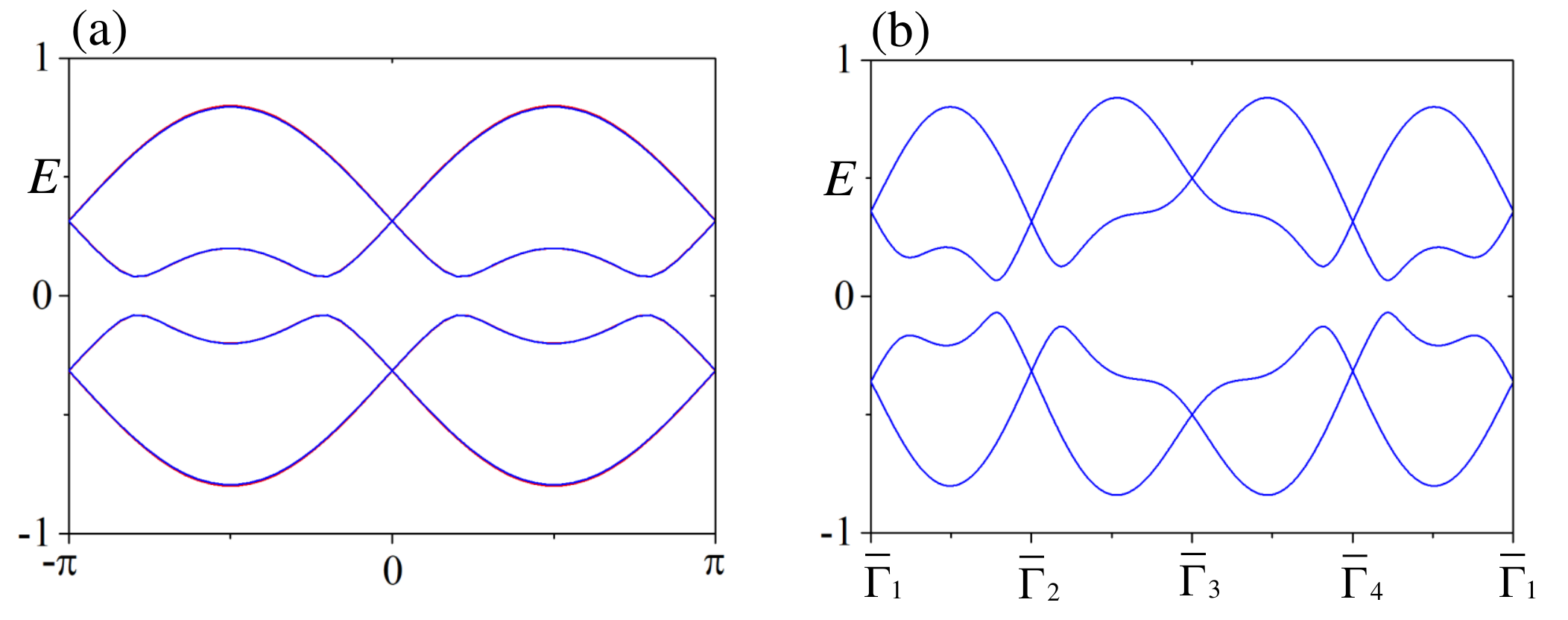}}
\caption{ Superconducting boundary bands. (a) The parameters of the 2D BdG Hamiltonian are $\epsilon_{1,2}=0$, $m_{0}=0$,
$m_{1}=1$, $m_{2}=2$, $\lambda_{1}=0.5$, $\lambda_{2}=1.0$, $\mu=0.3$, $\Delta_{0}=0$ and $\Delta_{1}=0.2$. The lattice size
along the direction with open boundary condition is $100$ lattice spacings. The numerical (blue curves)
and analytical (red curves) results show excellent agreement. (b) The parameters of the 3D BdG Hamiltonian are $\epsilon_{1,2,3}=0$, $m_{0}=0$,
$m_{1,2}=1$, $m_{3}=3$, $\lambda_{1,2}=0.5$, $\lambda_{3}=1.5$, $\mu=0.3$, $\Delta_{0}=0.1$ and $\Delta_{1}=0.15$. The lattice size
along the direction with open boundary condition is $100$ lattice spacings. The numerical (blue curves)
and analytical (red curves) results show perfect agreement.
}  \label{match2}
\end{figure}

\end{widetext}

\end{document}